\documentclass[aps,preprint]{revtex4}
\usepackage{amsmath}
\usepackage[english]{babel}
\usepackage[latin1]{inputenc}
\usepackage{amssymb}
\usepackage{graphicx}
\usepackage{natbib}
\usepackage{epsfig}
\usepackage{bm}
\usepackage{hyperref}
\usepackage{dcolumn}
\usepackage{graphicx}
\usepackage{color}
\begin{document}

\title{\sc\Large{Pion-to-vacuum vector and axial vector amplitudes and weak decays of pions in a magnetic field}}

\author{M. Coppola$^{a,b}$, D. Gomez Dumm$^{c}$, S. Noguera$^{d}$ and N.N.\ Scoccola$^{a,b,e}$}

\affiliation{$^{a}$ CONICET, Rivadavia 1917, 1033 Buenos Aires, Argentina}
\affiliation{$^{b}$ Physics Department, Comisi\'{o}n Nacional de Energ\'{\i}a At\'{o}mica, }
\affiliation{Avenue Libertador 8250, 1429 Buenos Aires, Argentina}
\affiliation{$^{c}$ IFLP, CONICET $-$ Departamento de F\'{\i}sica, Facultad de Ciencias Exactas,
Universidad Nacional de La Plata, C.C. 67, 1900 La Plata, Argentina}
\affiliation{$^{d}$ Departamento de F\'{\i}sica Te\'{o}rica and IFIC, Centro Mixto
Universidad de Valencia-CSIC, E-46100 Burjassot (Valencia), Spain}
\affiliation{$^{e}$ Universidad Favaloro, Sol{\'{\i}}s 453, 1078 Buenos Aires, Argentina
\vspace*{2cm}}

\begin{abstract}
We propose a model-independent parametrization for the one-pion-to-vacuum
matrix elements of the vector and axial vector hadronic currents in the
presence of an external uniform magnetic field. It is shown that, in
general, these hadronic matrix elements can be written in terms of several
gauge covariant Lorentz structures and form factors. Within this framework
we obtain a general expression for the weak decay $\pi^- \to l\,\bar\nu_l$
and discuss the corresponding limits of strong and weak external magnetic
fields.
\end{abstract}

\pacs{}

\maketitle

\renewcommand{\thefootnote}{\arabic{footnote}}
\setcounter{footnote}{0}

\section{Introduction}

The effect of intense magnetic fields on the properties of strongly
interacting matter has gained significant interest in recent years
\cite{Kharzeev:2012ph,Andersen:2014xxa,Miransky:2015ava}. This is mostly
motivated by the realization that strong magnetic fields might play an
important role in the study of the early Universe~\cite{Grasso:2000wj}, in
the analysis of high energy noncentral heavy ion collisions~\cite{HIC}, and
in the description of physical systems such as compact stellar objects
called magnetars~\cite{duncan}. It is well known that magnetic fields also
induce interesting phenomena like the enhancement of the QCD vacuum (the
so-called ``magnetic catalysis'')~\cite{Gusynin:1994re}, and the decrease of
critical temperatures for chiral restoration and deconfinement QCD
transitions~\cite{Bali:2011qj}.

In this work we concentrate on the effect of a magnetic field
$\vec B$ on the weak pion-to-lepton decay $\pi^-\to l\,\bar\nu_l$.
In fact, the study of weak decays of hadrons in the presence of
strong electromagnetic fields has a rather long history. Already
in the mid 1960s, the effect of some particularly simple
electromagnetic field configurations on leptonic decays of charged
pions was considered~\cite{Nikishov:1964zza,Nikishov:1964zz}. Some
years later, the decay of a neutron in the presence of a magnetic
field was also studied~\cite{Matese:1969zz,FassioCanuto:1970wk}.
In most of the existing calculations of the decay rates, however,
the effect of the external field on the internal structure of the
participating particles has not been taken into account. Only in
recent years has the $B$-dependence of pion masses actively been 
investigated from the theoretical point of view. This has been
done using approaches like e.g.~chiral perturbation
theory~\cite{Agasian:2001ym,Andersen:2012zc}, quark effective
models~\cite{Fayazbakhsh:2012vr,Avancini:2015ady,Zhang:2016qrl,Avancini:2016fgq,Mao:2017wmq,
GomezDumm:2017jij,Andreichikov:2018wrc,Orlovsky:2013wjd,Andreichikov:2016ayj,Wang:2017vtn,Liu:2018zag,Coppola:2018vkw}
and lattice QCD (LQCD)
simulations~\cite{Bali:2011qj,Luschevskaya:2014lga,Bali:2017ian}.
In addition, some of these
works~\cite{Agasian:2001ym,Andersen:2012zc,Fayazbakhsh:2012vr,Avancini:2015ady,Zhang:2016qrl,Avancini:2016fgq,Mao:2017wmq,GomezDumm:2017jij,Andreichikov:2018wrc}
(see also Refs.~\cite{Fayazbakhsh:2013cha,Simonov:2015xta})
considered, for the case of the neutral pion, the $B$-dependence of
the decay constant $f_{\pi^0}$. For the charged pion, such an
effect has been analyzed in the context of chiral perturbation
theory~\cite{Andersen:2012zc}, quark-antiquark effective chiral
models~\cite{Simonov:2015xta} and very recently through LQCD
calculations~\cite{Bali:2018sey}. An interesting observation was
made in Ref.~\cite{Fayazbakhsh:2013cha}, where it is claimed that,
due to the explicit breaking of rotational invariance caused by
the magnetic field, one can define two different decay constants,
one of them associated with the direction parallel to $\vec B$ and
another one associated with the perpendicular directions. A
further relevant statement has been pointed out in
Ref.~\cite{Bali:2018sey}. In that work it is noted that the
existence of the background field opens the possibility of a
nonzero pion-to-vacuum transition via the vector piece of the
electroweak current, which implies the existence of a further
decay constant. Taking into account this new constant, the authors
of Ref.~\cite{Bali:2018sey} obtained an expression for the decay
width under the assumption that the decaying pion is at rest.
However, one should note that in the case of a charged pion this
condition cannot be exactly fulfilled, due to the presence of the
magnetic field. Even in its lowest energy level (hence, in the
lowest Landau level), the charged pion still keeps a zero point
motion. In this context, the aim of the present work is twofold.
Firstly, considering an external uniform magnetic field, we
determine the form of the most general model-independent hadronic
matrix elements, written in terms of gauge covariant Lorentz
structures. In particular, we show that in the presence of a
magnetic field several independent form factors can be in
principle defined. Secondly, given the general form of the
hadronic matrix elements, we obtain an expression for the
$\pi^-\to l\,\bar\nu_l$ decay width, taking into account the
effect of the magnetic field on both pion and lepton
wave functions.

This paper is organized as follows. In Sec.~II we discuss the structure of
the pion-to-vacuum matrix elements in the presence of a uniform static
magnetic field from a general point of view. We start by identifying the
relevant gauge covariant Lorentz structures, and then we proceed to obtain
the hadronic matrix elements for neutral and charged pions. In Sec.~III we
obtain an explicit expression for the $\pi^\pm$ leptonic weak decay width.
After considering the general case of an arbitrary magnetic field strength,
we discuss the particular cases of both strong and weak magnetic fields.
Finally, in Sec.~IV we present our main conclusions. We also include
Appendixes A, B, C and D to quote technical details of our calculations.

\section{Pion-to-vacuum amplitudes in the presence of an external uniform magnetic field: general structure}

In this section we analyze the general form of the pion-to-vacuum matrix
elements of vector and axial vector quark currents. Throughout this work we
use the Minkowski metric $g^{\mu\nu} = \mbox{diag}(1,-1,-1,-1)$, as well as
the convention $\epsilon^{0123}= +1$ for the totally antisymmetric tensor
$\epsilon^{\mu\nu\alpha\beta}$.

Let us start by considering the hadronic matrix elements for the case of a
neutral pion in the absence of the external magnetic field. The amplitudes for
the vector and axial vector quark currents are
\begin{eqnarray}
H^{0,\mu}_V(x,\vec p) &=& \langle 0 | \bar \psi(x) \gamma^\mu \frac{\tau_3}{2}\,\psi(x) | \pi^0(\vec p\,) \rangle\ , \nonumber \\
H^{0,\mu}_A(x,\vec p) &=& \langle 0 | \bar \psi(x) \gamma^\mu \gamma_5 \frac{\tau_3}{2}\,\psi(x) | \pi^0(\vec p\,)
\rangle\ ,
\label{had0}
\end{eqnarray}
where $\psi(x)$ includes the $u$ and $d$ quark fields,
\begin{equation}
\psi\left(x\right)=\left(\begin{array}{c}
\psi_{u}(x)\\
\psi_{d}(x)
\end{array}\right)\ ,
\end{equation}
and $\tau_3$ is a Pauli matrix that acts on flavor space. To deal with the
matrix elements in Eqs.~(\ref{had0}) it is possible to hadronize the quark
currents, i.e., to consider matrix elements of hadronic field operators
carrying appropriate Lorentz indices and quantum numbers. In the low energy
limit (typically, below the $\rho$ meson threshold), the relevant hadronic
field is the pion field $\phi_{\pi^0}(x)$, and (in absence of external
fields) the only available vectorlike differential operator is the momentum
operator $\hat p^\mu = i\partial^\mu$. Since the pion field is pseudoscalar,
only the matrix element of the axial-vector hadronic current can be nonzero.
In this way, one has
\begin{eqnarray}
\langle 0 | \bar{\psi}(x)\gamma^{\mu}\frac{\tau_3}{2}\psi(x)  | \pi^0(\vec p\,) \rangle & = & 0\ ,
\nonumber \\
\langle 0 |
\bar{\psi}(x)\gamma^{\mu}\gamma_5\frac{\tau_3}{2}\psi(x) | \pi^0(\vec p\,) \rangle
& = & f(\hat p^2)\,  \partial^{\mu} \langle 0 |\phi_{\pi^0} (x) | \pi^0(\vec p\,) \rangle \ .
\end{eqnarray}
Here, the function $f(\hat p^2)$ contains all the information of
nonperturbative QCD contributions. Using the explicit form of
$\phi_{\pi^0}(x)$ and the commutation rules for
the corresponding creation and annihilation operators [see
Eqs.~(\ref{bosonfield}) and (\ref{commbpart})], one immediately
finds
\begin{eqnarray}
H^{0,\mu}_V(x,\vec p\,) &=& 0\ , \nonumber \\
H^{0,\mu}_A(x,\vec p\,) &=& - i f(p^2)\, p^\mu e^{-i p\cdot x}\ .
\end{eqnarray}
As usual, the four-momentum $p^\mu$ is defined by $p^\mu = (E_\pi, \vec p\,)$,
with $E_\pi = \sqrt{m_\pi^2 + |\vec p\,|^2}$. Similar expressions can be
obtained for charged pions. It can be seen that the invariance of
strong interactions under ${\cal P}$, ${\cal C}$ and ${\cal T}$
transformations implies that $f(p^2)$ is a real function. In the absence of
external fields, the pion decay constant is given by $f_\pi =
f(m_\pi^2)\simeq 92.3$ MeV~\cite{Tanabashi:2018oca}.

We turn now to the situation in which a static external electromagnetic
field is present. In this case, other tensor structures are possible. For a
particle of charge $Q$ the relevant basic tensors are the gauge covariant
derivative ${\cal D}^{\mu}$ and the gauge invariant electromagnetic
field strength $F^{\mu\nu}$, defined as
\begin{equation}
{\cal D}^{\mu} \ = \ \partial^{\mu} + i\, Q\, {\cal A}^{\mu} \ , \qquad
F^{\mu\nu} \ = \ \partial^{\mu}{\cal A}^{\nu}-\partial^{\nu}{\cal A}^{\mu} \ .
\end{equation}
Taking them as building blocks, one can in principle obtain an
infinite number of differential operators with different Lorentz tensor
structures. However, for the particular case of a uniform static magnetic
field $\vec B$, it is well known (see e.g.~Ref.~\cite{Dittrich:2000zu}) that
only a few independent tensors exist. Noting that $F^{0i} =0$ and $F^{ij} =
F_{ij} = - \epsilon_{ijk} B^k$, we get
\begin{equation}
\left[{\cal D}^{\mu},{\cal D}^{\nu}\right] \ = \ iQF^{\mu\nu} \ = \
-iQ \, \epsilon^{0\mu\nu k} \, B^k\ ,  \qquad \qquad  k=1,2,3\ .
\end{equation}
For definiteness, and without losing generality, in what follows we take
$B^k = B \, \delta_{k3}$. Using the above relations it is easy to prove that
one has only two independent scalars, apart from the particle electric charge $Q$
and $F^{\mu\nu}F_{\mu\nu}=2B^2$. These can be taken to be
\begin{equation}
{\cal D}_{||}^2 \ = \ ({\cal D}^3)^2 - ({\cal D}^0)^2 \ , \qquad {\cal
D}_\perp^2 \ = \ -({\cal D}^1)^2 - ({\cal D}^2)^2\ .
\end{equation}
In addition, it is possible to find four independent four-vectors. One
possible choice is the set
\begin{eqnarray}
{\cal D}^{\mu} &=& ({\cal D}^0, \vec {\cal D}) \ , \nonumber \\
-i\, F^{\mu\nu}{\cal D}_{\nu} &=& -i\, B \ (0,{\cal D}^2,-{\cal D}^1,0) \ , \nonumber \\
F^{\mu\nu}F_{\nu\alpha}{\cal D}^{\alpha} &=& - B^2 \ (0,{\cal D}^1,{\cal D}^2,0) \ , \nonumber \\
\dfrac{1}{2} \ \epsilon^{\mu\nu\alpha\beta}F_{\nu\alpha}{\cal D}_{\beta} &=& B \ ({\cal D}^3,0,0,{\cal D}^0)
\label{vecop}
\end{eqnarray}
(notice that the last of these tensors transforms in fact as an axial vector).

From the above expressions for Lorentz scalars and four-vectors, we can
write now a general form for the hadronic currents we are interested in. We
consider first the case of the neutral pion, for which $Q=0$ and the
operator ${\cal D}^\mu$ reduces to the usual derivative $\partial^\mu$.
Taking into account once again the intrinsic parity of the pion field, one
has
\begin{eqnarray}
\langle 0 | \bar{\psi}(x)\gamma^{\mu}\frac{\tau_3}{2}\psi(x)  | \pi^0(\vec p\,) \rangle
& = &
\hat f_{\pi^0}^{(V)}\; (\partial^3,0,0,\partial^0)\;  \langle 0 |\phi_{\pi^0} (x) | \pi^0(\vec p\,) \rangle \ ,
\label{fv0} \\
\langle 0 |
\bar{\psi}(x)\gamma^{\mu}\gamma_5\frac{\tau_3}{2}\psi(x) | \pi^0(\vec p\,) \rangle
& = &
\Big[\hat f_{\pi^0}^{(A1)}\,
(   \partial^0, \vec \partial)  \, -i \hat f_{\pi^0}^{(A2)} \, (0,\partial^2,-\partial^1,0)
 \
\nonumber\\
& & -  \, \hat f_{\pi^0}^{(A3)} \,  (0,\partial^1,\partial^2,0) \,   \Big]
\langle 0 |\phi_{\pi^0} (x) | \pi^0(\vec p\,) \rangle \ ,
\label{fa0}
\end{eqnarray}
where $\hat f_{\pi^0}^{(V)}$ and $\hat f_{\pi^0}^{(Ai)}$ are complex functions of the magnetic field
and the scalar differential operators $\partial^2_{||}$ and
$\partial^2_{\perp}$. The hadronic matrix elements can be readily obtained
using Eqs.~(\ref{bosonfield}) and (\ref{commbpart}). We find it convenient
to define some linear combinations of the respective Lorentz components,
namely
\begin{eqnarray}
H_{\parallel,V}^{0,\,\epsilon} (x,\vec p\,) & \equiv &
H_{V}^{0,0} (x,\vec p\,) + \epsilon \, H_{V}^{0,3}(x,\vec p\,) =  -i\,\epsilon\, f_{\pi^0}^{(V)} \left( E_{\pi^0} + \epsilon\, p^3 \right) \,
e^{-i p\cdot x}\ ,
\label{fv01} \\
H_{\perp,V}^{0,\,\epsilon}(x,\vec p\,) & \equiv & H_{V}^{0,1}(x,\vec p\,) + i \epsilon \,
H_{V}^{0,2}(x,\vec p\,) = 0 \ ,
\label{fv02}
\end{eqnarray}
and
\begin{eqnarray}
H_{\parallel,A}^{0,\,\epsilon}(x,\vec p\,) & \equiv &
H_{A}^{0,0}(x,\vec p\,)  + \epsilon\, H_{A}^{0,3}(x,\vec p\,) \nonumber \\
& = & -\, i f_{\pi^0}^{(A1)}
( E_{\pi^0} + \epsilon\,  p^3 )\, e^{-i p\cdot x}\ ,
\label{fa01} \\
H_{\perp,A}^{0,\,\epsilon}(x,\vec p\,) & \equiv & H_{A}^{0,1}(x,\vec p\,) + i\epsilon\,
H_{A}^{0,2}(x,\vec p\,) \nonumber \\
& = & - i \left[ f_{\pi^0}^{(A1)} -\epsilon f_{\pi^0}^{(A2)} -
f_{\pi^0}^{(A3)}\right]  (p^1 + i\epsilon  p^2)\,  e^{-i p\cdot x}\ ,
\label{fa02}
\end{eqnarray}
with $\epsilon = \pm$. Here, $f_{\pi^0}^{(V)}$ and $f_{\pi^0}^{(Ai)}$ are
functions of $p^2_{\perp}=(p^1)^2+(p^2)^2$ and
$p^2_\parallel=E_{\pi^0}^2-(p^3)^2$, with $p_\parallel^2-p_\perp^2 = p^2 =
m_{\pi^0}^2$. Notice that, although it is not indicated explicitly, the pion 
mass $m_{\pi^0}$ is a function of the magnetic field
$B$. As in the $B=0$ case, it is important to consider the constraints on
the form factors arising from the discrete symmetries of the interaction
Lagrangian in the presence of the magnetic field. This is discussed in some
detail in Appendix B, where it is shown that these symmetries lead to
$f_{\pi^0}^{(A2)}=0$ while the remaining form factors turn out to be real.
In this way, we conclude that the most general forms of the vector and axial
vector pion-to-vacuum matrix elements, in the presence of an external
constant and homogenous magnetic field along the 3-axis, are
\begin{eqnarray}
H^{0,\,\epsilon}_{\parallel,V} (x,\vec p\,) & = &  -i\epsilon\, f_{\pi^0}^{(V)}
\left( E_{\pi^0} + \epsilon\; p^3 \right) \, e^{-i p\cdot x}\ ,
\label{fv01f} \\
H^{0,\,\epsilon}_{\perp,V}(x,\vec p\,) &=& 0
\label{fv02f}
\end{eqnarray}
and
\begin{eqnarray}
H^{0,\,\epsilon}_{\parallel,A}(x,\vec p\,) &=& -\, i f_{\pi^0}^{(A1)}
\left( E_{\pi^0} + \epsilon\,  p^3 \right) \, e^{-i p\cdot x}\ ,
\label{fa01f} \\
H^{0,\,\epsilon}_{\perp,A}(x,\vec p\,) & = & -\, i \left[ f_{\pi^0}^{(A1)}
- f_{\pi^0}^{(A3)}\right]
\left( p^1 + i\epsilon\,  p^2  \right)  e^{-i p\cdot x}\ ,
\label{fa02f}
\end{eqnarray}
where all form factors are real. The results in
Eqs.~(\ref{fa01f}) and (\ref{fa02f}) are in agreement with the observation made in
Ref.~\cite{Fayazbakhsh:2013cha} that, due to the explicit breaking of rotational invariance caused by
the magnetic field, one can define for the neutral meson two different form
factors related to the axial current. One of them can be associated with the
direction parallel to $\vec B$, and the other one with the perpendicular
directions. In addition, according to Eq.~(\ref{fv01f}) we find that a
further form factor related to the vector current can be defined as well.

Let us consider now the case of the charged pion $\pi^\sigma$, with
$\sigma=\pm\,$ (electric charge $Q=\sigma |e|$). We adopt in this work the
Landau gauge, in which ${\cal A}^\mu = (0, 0, B x^1, 0)$. Then, ${\cal
D}^\mu =
\partial^\mu + i s B_e \, x^1 \delta_{\mu 2}$, with $s = \mbox{sign}(\sigma
B)$ and $B_e= |e B|$.
One has in this case
\begin{eqnarray}
H_{V}^{\sigma,\mu} (x,\breve p) & = &\langle 0 |
\bar{\psi}(x) \gamma^{\mu}\tau^{-\sigma} \psi(x) | \pi^\sigma(\breve p) \rangle
\nonumber \\
& = & \sqrt{2}\, ({\cal D}^3,0,0,{\cal D}^0)\, \hat f_{\pi^\sigma}^{(V)}\,  \langle 0 |
\phi^s_{\pi^\sigma}(x) | \pi^\sigma(\breve p) \rangle\
\label{hvch}
\end{eqnarray}
and
\begin{eqnarray}
H_{A}^{\sigma,\mu} (x,\breve p) & = & \langle 0 | \bar{\psi}(x)\gamma^{\mu}\gamma_5
\tau^{-\sigma} \psi(x) | \pi^\sigma(\breve p) \rangle \nonumber \\
& = & \sqrt{2}\, \Big[
  ({\cal D}^0, \vec {\cal D})\,
\hat f_{\pi^\sigma}^{(A1)} \, -is \, (0,{\cal D}^2,-{\cal D}^1,0)\, \hat f_{\pi^\sigma}^{(A2)}
\, - \, (0,{\cal D}^1,{\cal D}^2,0)\, \hat f_{\pi^\sigma}^{(A3)} \
\Big] \times \nonumber \\
& & \langle 0 | \phi^s_{\pi^\sigma} (x)  | \pi^\sigma(\breve p) \rangle \ ,
\label{hva}
\end{eqnarray}
where $\tau^\pm = (\tau_1 \pm i \tau_2)/2$. Here, $\hat
f_{\pi^\sigma}^{(V)}$ and $\hat f_{\pi^\sigma}^{(Ai)}$ are functions of the
scalar operators ${\cal D}^2_{||}$ and ${\cal D}^2_{\perp}$, while $\breve p
= (\ell, p^2,p^3)$, $\ell$ being a non-negative integer (Landau index),
denotes a set of quantum numbers that characterize the charged pion state in
the presence of the magnetic field (see Appendix A.2 for details). The
sign $s$ in the term carrying $\hat f_{\pi^\sigma}^{(A2)}$ has been
conventionally introduced for later convenience. Notice that in
Eqs.~(\ref{hvch}) and (\ref{hva}) the operators $\hat f_{\pi^\sigma}^{(V)}$
and $\hat f_{\pi^\sigma}^{(Ai)}$ have been placed to the right of the
operators listed in Eq.~(\ref{vecop}). In fact, the ordering is relevant,
since (contrary to the case of the neutral pion) the operators ${\cal D}^1$
and ${\cal D}^2$ do not commute. On the other hand, since the basis formed
by the operators in Eq.~(\ref{vecop}) is complete, our ordering choice does
not imply a loss of generality (different orderings just correspond to
alternative definitions of the form factors). The convenience of our
election will become clear below.

To proceed we need the general form of the $\pi^\sigma$ wave functions in the
presence of the external magnetic field. For our choice of gauge and
magnetic field direction these are given by Eq.~(\ref{chargepionexp}). From
this equation, together with the commutation relations of the associated
creation and annihilation operators given in Eq.~(\ref{conbos}), one can
obtain general expressions for the hadronic matrix elements. As in the case
of the neutral pion, these can be conveniently written in terms of linear
combinations of their Lorentz components. We have
\begin{eqnarray}
H^{\sigma,\,\epsilon}_{\parallel,V} (x,\breve p) & \equiv &
H_V^{\sigma,0} (x,\breve p) + \epsilon\, H_V^{\sigma,3}(x,\breve p) \;
= \; \epsilon\,\sqrt{2}\ {\cal D}_\parallel^\epsilon  \; \hat f_{\pi^\sigma}^{(V)}
\; \mathbb{F}^s_{\bar p}(x)
\ ,\nonumber \\
H^{\sigma,\,\epsilon}_{\perp,V}(x,\breve p) & \equiv &
H_V^{\sigma,1}(x,\breve p)+ i\, \epsilon\, H_V^{\sigma,2}(x,\breve p) \; = \; 0
\label{hsv}
\end{eqnarray}
and
\begin{eqnarray}
\hspace{-0.5cm}
H^{\sigma,\,\epsilon}_{\parallel,A}(x,\breve p) & \equiv & H_A^{\sigma,0}(x,\breve p) +\epsilon\, H_A^{\sigma,3}(x,\breve p) \;
= \;\sqrt{2} \ {\cal D}_\parallel^\epsilon \; \hat f_{\pi^\sigma}^{(A1)} \; \mathbb{F}^s_{\bar p}(x) \ ,
\nonumber \\
\hspace{-0.5cm}
H^{\sigma,\,\epsilon}_{\perp,A}(x,\breve p) & \equiv &
H_A^{\sigma,1}(x,\breve p) + i \, \epsilon\  H_A^{\sigma,2}(x,\breve p) \, =
\,\sqrt{2} \ {\cal D}_\perp^\epsilon \, \big(\hat f_{\pi^\sigma}^{(A1)} -
s\,\epsilon \, \hat f_{\pi^\sigma}^{(A2)}  - \hat
f_{\pi^\sigma}^{(A3)}\big)\;
 \mathbb{F}^s_{\bar p}(x) \ ,
\label{hsa}
\end{eqnarray}
where ${\cal D}_\parallel^\epsilon = {\cal D}^0 +\epsilon\, {\cal D}^3$ and
${\cal D}_\perp^\epsilon = {\cal D}^1 +i\epsilon {\cal D}^2$. We use here
the notation $\bar p = (E_{\pi^\sigma} , \breve p)$, with $E_{\pi^\sigma} =
\sqrt{m_{\pi^\sigma}^2 + (2\ell+1)B_e +(p^3)^2}$ (in fact, one has
$E_{\pi^+}=E_{\pi^-}$, the charge index $\sigma$ being kept in order to
distinguish $E_{\pi^\pm}$ from $E_{\pi^0}$). The functions
$\mathbb{F}^{\pm}_{\bar p}(x)$ are eigenfunctions of the charged pion
Klein-Gordon equation in the presence of the magnetic field, for our choices
of gauge and magnetic field direction. Their explicit expressions are given
in Eq.~(\ref{efes}). Notice that on the rhs of Eqs.~(\ref{hsv}) and
(\ref{hsa}) we have differential operators acting on these functions. Using
the relations
\begin{eqnarray}
{\cal D}_\parallel^\epsilon\,\mathbb{F}^s_{\bar p}(x) & = &
-i\, (E_{\pi^\sigma} + \epsilon\, p^3) \;  \mathbb{F}^s_{\bar p}(x)\ ,
 \\
{\cal D}_\perp^\epsilon\; \mathbb{F}^s_{\bar p}(x) & = &
-s\,\epsilon \sqrt{B_e \left(2 \ell+1 -s\,\epsilon \right)}\;\mathbb{F}^s_{\bar p -s\,\epsilon}(x)
\end{eqnarray}
and
\begin{eqnarray}
{\cal D}^2_{||}\;\mathbb{F}^s_{\bar p}(x) & = &
\Big[ E_{\pi^\sigma}^2 -(p^3)^2\Big] \ \mathbb{F}^s_{\bar p}(x)\ ,
\label{eq3} \\
{\cal D}^2_{\perp}\;\mathbb{F}^s_{\bar p}(x) & = &
\left(2 \ell +1\right) B_e \ \mathbb{F}^s_{\bar p}(x)\ ,
\label{eq4}
\end{eqnarray}
where $\bar p\pm 1 \equiv (E_{\pi^\sigma},\breve p\pm 1)$, with $\breve p\pm 1 = (\ell\pm
1,p^2,p^3)$, we finally obtain
\begin{eqnarray}
H^{\sigma,\,\epsilon}_{\parallel,V} (x,\breve p) & = &
- i \epsilon\,\sqrt{2} \, f_{\pi^\sigma}^{(V)} \, (E_{\pi^\sigma} + \epsilon\, p^3) \ \mathbb{F}^s_{\bar p}(x) \ ,
\label{hvpar} \\
H^{\sigma,\,\epsilon}_{\perp,V} (x,\breve p) & = & 0
\label{hvv}
\end{eqnarray}
and
\begin{eqnarray}
H^{\sigma,\,\epsilon}_{\parallel,A} (x,\breve p) & = &
-\, i\, \sqrt{2}\,  f_{\pi^\sigma}^{(A1)} \, (E_{\pi^\sigma} + \epsilon \,p^3) \ \mathbb{F}^s_{\bar p}(x)\ ,
\label{hapar} \\
H^{\sigma,\,\epsilon}_{\perp,A} (x,\breve p) & = &
-s\,\epsilon\,\sqrt{2}\, \big(f_{\pi^\sigma}^{(A1)}  -
s\,\epsilon \, f_{\pi^\sigma}^{(A2)}
-f_{\pi^\sigma}^{(A3)} \big) \sqrt{B_e\left( 2 \ell+ 1
-s\,\epsilon \right)} \ \mathbb{F}^s_{\bar p -s\,\epsilon}(x)\ .
\label{hav}
\end{eqnarray}
In the above expressions, the form factors $f_{\pi^\sigma}^{(V)}$,
$f_{\pi^\sigma}^{(Ai)}$ arise from the action of the operators $\hat
f_{\pi^\sigma}^{(V)}$, $\hat f_{\pi^\sigma}^{(Ai)}$ on the functions
$\mathbb{F}^s_{\bar p}(x)$. These operators are, in general, gauge-dependent
functions of the operators $\mathcal{D}_{\Vert}^{2}$ and
$\mathcal{D}_{\bot}^{2}$. However, for an on-energy-shell pion, taking into
account Eqs.~(\ref{eq3}) and (\ref{eq4}) it can be seen that the resulting
form factors $f_{\pi^{\sigma}}^{\left(V\right)}$ and
$f_{\pi^{\sigma}}^{\left(Ai\right)}$ turn out to be gauge-independent
functions of the pion mass, $m_{\pi^{\sigma}}$; the third component of the
momentum, $p^{3}$; the Landau index, $\ell$; and the magnetic field, $B$
(both explicitly and through the charged pion mass). In the so-called
symmetric gauge, Eqs.~(\ref{eq3}) and (\ref{eq4}) should be modified by
changing the functions $\mathbb{F}_{\bar{p}}^{\sigma}\left(x\right)$ by the
corresponding ones in that gauge, which involve associated Laguerre
polynomials~\cite{Wakamatsu:2017isl}. On the other hand, notice that the
eigenvalues in Eqs.~(\ref{eq3}) and (\ref{eq4}) are functions of $|B|$,
while the scalar combination $F^{\mu\nu}F_{\mu\nu}$ depends on $B^2$.
Therefore, taking into account the form of the four-vectors in
Eq.~(\ref{vecop}), and the factor $s$ introduced in the term carrying $\hat
f_{\pi^\sigma}^{(A2)}$ in Eq.~(\ref{hva}), it is seen that the vector form
factor $f_{\pi^\sigma}^{(V)}$ should be odd under the exchange $B \to -B$,
while the axial vector form factors $f_{\pi^\sigma}^{(Ai)}$ should be even
functions of $B$. For definiteness, in what follows we will take $B>0$, and
consequently $s=\sigma$.

As in the case of the neutral pion, the discrete symmetries of the
interaction Lagrangian in the presence of the magnetic field lead to some
restrictions on the form factors $f_{\pi^\sigma}^{(V)}$ and
$f_{\pi^\sigma}^{(Ai)}$. Indeed, as shown in Appendix B, they have to be real
and independent of the sign of the pion charge. Finally, let us point out
that the form factors $f_{\pi^\sigma}^{(A1)}$ appearing in
Eqs.~(\ref{hapar}) and (\ref{hav}) are the same. In fact, this is a
consequence of our choice of ordering in Eq.~(\ref{hva}). Had we put the
operator functions $\hat f_{\pi^\sigma}^{(Ai)}$ to the left of the other
operators in Eq.~(\ref{hva}), the form factor $f_{\pi^\sigma}^{(A1)}$ in
Eq.~(\ref{hapar}) would have arisen from the action of $\hat
f_{\pi^\sigma}^{(A1)}$ on $\mathbb{F}^\sigma_{\bar p}(x)$, while the one in
Eq.~(\ref{hav}) would correspond to the action of $\hat
f_{\pi^\sigma}^{(A1)}$ on $\mathbb{F}^\sigma_{\bar p -\sigma\epsilon}(x)$.
In any case, it should be stressed that this would also imply a redefinition
of $f_{\pi^\sigma}^{(A2)}$ and $f_{\pi^\sigma}^{(A3)}$, in such a way that
the contribution of $H^{\sigma,\,\epsilon}_{\perp,A} (x,\breve p)$ to any
physical quantity remains unchanged.

\section{Weak decay width of charged pions under a magnetic field }

\subsection{Basic relations and decay kinematics}

Let us analyze the decay width for the process $\pi^-\to l\,\bar\nu_l$, with
$l = \mu, e$, in the presence of the external magnetic field. Following the
notation introduced in the previous section, the initial charged pion state
is determined by the quantum numbers $\breve p = (\ell, p^2,p^3)$, the
associated energy being $E_{\pi^-} = \sqrt{m_{\pi^-}^2+(2\ell+1)B_e +
(p^3)^2}$. The quantum numbers corresponding to the outgoing lepton state
are taken to be $\breve q = (n,q^2,q^3)$, together with a polarization index
$r_l$. In this case the energy is given by $E_l = \sqrt{m_l^2+2nB_e +
(q^3)^2}$ (see Appendix A.3). Finally, being electrically neutral, the outgoing
antineutrino is taken to be in a state of momentum $\vec k = (k^1,k^2,k^3)$
and polarization $r_{\bar\nu_l}$, with energy $E_{\bar\nu_l} =
\sqrt{(k^1)^2+(k^2)^2+(k^3)^2} = |\vec k|$.

On general grounds, the decay width for the process is given by
\begin{eqnarray}
\Gamma_l^-(B) &=& \lim_{T\rightarrow\infty} \dfrac{1}{n_{\pi^-}} \sum_{r_l}
\sum_{n=0}^\infty\;\int \frac{dq^2\,dq^3}{(2\pi)^3 2E_l}  \sum_{r_{\bar\nu_l}} \int
\dfrac{d^3k}{(2\pi)^3 2E_{\bar\nu_l}} \,\dfrac{|({\mathcal S}-1)_{fi}|^2}{T}\ ,
\label{decB}
\end{eqnarray}
where $({\mathcal S}-1)_{fi}$ is the relevant ${\mathcal S}$-matrix element
between the initial and final states, and the particle number $n_{\pi^-}$
associated with the initial $\pi^-$ state is given in Eq.~(\ref{numberb}).
Thus, one has
\begin{equation}
\Gamma_l^-(B)\ = \ \lim_{S,\,T\rightarrow\infty} \dfrac{1}{2 E_{\pi^-}}
\sum_{n=0}^\infty\,\int
\dfrac{dq^2\,dq^3}{(2\pi)^3 2E_l}\, \dfrac{d^3k}{(2\pi)^3 2E_{\bar\nu_l}} \sum_{r_l,\,r_{\bar\nu_l}}
\frac{|\langle\, l(\breve q , r_l)\, \bar \nu_l( \vec k, r_{\bar\nu_l} ) | \mathcal{L}_W |
\pi^-(\breve p)\,\rangle |^2}{2\pi S\, T}\ ,
\label{uno}
\end{equation}
where $\mathcal{L}_W$ is the usual $V-A$ four-fermion effective weak
interaction Langrangian integrated over space-time, while $T$ and $S$ 
are the time interval and the surface on the
$x^2x^3$ plane in which the interaction is active. At the end of the
calculation, the limit $S,T\to \infty$ will be taken. Let us recall that,
according to our gauge choice, the motion in the $x^1$ axis is bounded.
Using the notation introduced in Appendix A, the matrix element in
Eq.~(\ref{uno}) is given by
\begin{eqnarray}
\langle\, l(\breve q , r_l)\, \bar \nu_l(\vec k, r_{\bar\nu_l} ) | \mathcal{L}_W |
\pi^-(\breve p)\,\rangle & = &  - \frac{G_F}{\sqrt{2}}\, \cos \theta_c
\nonumber \\
& & \hspace{-1.7cm} \times \int d^4x \, H^{-,\mu}_L(x,\breve p)
\ \bar  U^-_l(x,\breve {q}, r_l) \, \gamma_\mu \, (1-\gamma_5) \,
V_{\nu_l}(x,\vec k, r_{\bar\nu_l})\ ,
\end{eqnarray}
where $H^{-,\mu}_L(x,\breve p)$ stands for the matrix element of the hadronic
current [see Eqs.~(\ref{hvch}) and (\ref{hva})],
\begin{equation}
H^{-,\mu}_{L}(x,\breve p) \ = \
H_V^{-,\mu} (x,\breve p) - H_A^{-,\mu} (x,\breve p) \ = \
\langle 0| \bar \psi_u(x)\, \gamma^\mu (1-\gamma_5)\, \psi_d(x)| \pi^-(\breve p) \rangle\ .
\label{hadronicme}
\end{equation}
Thus, from Eqs.~(\ref{UVneut}), (\ref{UVlept}) and (\ref{ep}), the decay amplitude
is given by
\begin{eqnarray}
\langle\, l(\breve q , r_l)\, \bar \nu_l(\vec k, r_{\bar\nu_l} ) | \mathcal{L}_W |
\pi^-(\breve p)\,\rangle & = &  - \frac{G_F}{\sqrt{2}}\, \cos \theta_c
\sum_{\lambda=\pm} \bar u^-_l(\breve {q}, r_l)\, \Delta^\lambda\,
\gamma_\mu (1-\gamma_5)\, v_{\nu_l}(\vec k, r_{\bar\nu_l}) \nonumber \\
& & \times \int d^4x \, H^{-,\mu}_{L}(x,\breve p)\, E^-_{\bar q,\lambda}(x)^\ast\, e^{ikx}\
,
\label{haches}
\end{eqnarray}
where $k^\mu=(E_{\bar\nu_l}, \vec k)$ and $\bar q = (E_l,\breve q)$. Given our
gauge choice for the external magnetic field, the electromagnetic potential
depends only on the $x^1$ coordinate. Consequently, the momentum is
conserved in the 0, 2 and 3 directions, leading to the corresponding delta
functions when the space-time integration in Eq.~(\ref{haches}) is
performed. We obtain
\begin{equation}
\int d^4x \ H_{L}^{-,\mu}(x,\breve p)\, E^-_{\bar q,\lambda}(x)^\ast\, e^{ikx}
\ = \ (2\pi)^3\,\delta(E_{\pi^-}-E_l-E_{\bar\nu_l})\,\prod_{i=2,3}
\delta(p^i-q^i-k^i)\,M^{-,\mu}_{\lambda}\ ,
\label{integral}
\end{equation}
where the $M^{-,\mu}_{\lambda}$ functions will be explicitly given below.
In this way, the amplitude can be written as
\begin{equation}
\langle\, l(\breve q , r_l)\, \bar \nu_l(\vec k, r_{\bar\nu_l} ) | \mathcal{L}_W |
\pi^-(\breve p)\,\rangle \ = \ (2\pi)^3
\, \delta(E_{\pi^-}-E_l-E_{\bar\nu_l}) \,\prod_{i=2,3}\delta(p^i-q^i-k^i)\; {\cal M}_{\pi^-\to\,
l\,\bar\nu_l}\ ,
\label{unnos}
\end{equation}
where
\begin{equation}
{\cal M}_{\pi^-\to\, l\,\bar\nu_l} \ = \ - \frac{G_F}{\sqrt{2}} \cos \theta_c\,
\sum_{\lambda=\pm} \bar  u^-_l(\breve {q}, r_l)\, \Delta^\lambda\, \gamma_\mu (1-\gamma_5)
\,v_{\nu_l}(\vec k, r_{\bar\nu_l})\; M^{-,\mu}_{\lambda}\ .
\label{doss}
\end{equation}

Thus, replacing in Eq.~(\ref{uno}) and taking the limit of infinite space-time volume,
the decay width is given by
\begin{equation}
\Gamma_l^-(B) = \frac{1}{4\pi E_{\pi^-}}\sum_{n=0}^\infty\int
\frac{dq^2\,dq^3}{2E_l}\,
\frac{d^3k}{2 E_{\bar\nu_l}}\, \frac{1}{(2\pi)^3}
\, \delta(E_{\pi^-}-E_l-E_{\bar\nu_l})\,\prod_{i=2,3}\delta(p^i-q^i-k^i)\;
\overline{\big|{\cal M}_{\pi^-\to\, l\,\bar\nu_l}\big|^2}\ ,
\label{gamgen}
\end{equation}
where
\begin{equation}
\overline{\big|{\cal M}_{\pi^-\to\, l\,\bar\nu_l}\big|^2} \ \equiv \
\sum_{r_l,\, r_{\bar\nu_l}}\Big|{\cal M}_{\pi^-\to\, l\,\bar\nu_l}\Big|^2 \ .
\label{ampa}
\end{equation}
As customary, when putting the expression on the rhs of Eq.~(\ref{unnos})
into Eq.~(\ref{uno}) we have replaced
$\left[\delta\left(E_{\pi^{-}}-E_{l}-E_{\bar{\nu}_{l}}\right)\right]^{2}$ by
$\lim_{T\rightarrow\infty}\,T\,\delta\left(E_{\pi^{-}}-E_{l}-E_{\bar{\nu}_{l}}\right)$
and $\left[\Pi_{i=2,3}\,\delta\left(p^{i}-q^{i}-k^{i}\right)\right]^{2}$ by
$\lim_{S\rightarrow\infty}\,S\,\Pi_{i=2,3}\,\delta\left(p^{i}-q^{i}-k^{i}\right)$.

Now, as it is usually done, we concentrate on the situation in which the
decaying pion is in the state of lowest energy. This corresponds to the case
$\ell =0$ and $p^3 =0$. Moreover, as will be shown below, neither the pion
energy nor the decay width depend on the value of $p^2$. The expression
for the decay width can be worked out, leading to
\begin{equation}
\Gamma_{l}^{-}(B) \ = \ \frac{1}{16\pi E_{\pi^-}^2}
\sum_{n=0}^{n_{\rm max}}
\,\int\frac{d^{2}k_{\perp}}{(2\pi)^2}\,\int\frac{dk^{3}}{2\pi}\ \frac{1}{\rule{0cm}{0.4cm}\bar{k^3}}\,
\big[\delta(k^{3}-\bar{k^3})+\delta(k^{3}+\bar{k^3})\big]\,
\overline{\big|{\cal M}_{\pi^-\to\, l\,\bar\nu_l}\big|^2}\ ,
\label{gamgen-1}
\end{equation}
where we have used the definitions
\begin{eqnarray}
n_{\rm max} & = & \frac{m_{\pi^-}^{2}-m_{l}^{2}+B_e}{2B_{e}}\ ,
\label{nmax} \\
\vec k_\perp & = & (k^1,k^2)\ , \\
\bar{k^3} & = &
\frac{\sqrt{E_{\pi^-}^{4}-2E_{\pi^-}^{2}\left(m_{l}^{2}+2nB_{e}+k_{\perp}^{2}\right)
+\left(m_{l}^{2}+2nB_{e}-k_{\perp}^{2}\right)^{2}}}{2E_{\pi^-}}\ ,
\label{k3}
\end{eqnarray}
and the integral over $k_\perp$ space is restricted to $|\vec k_{\perp}|
\leq E_{\pi^-}-\sqrt{m_l^2+2nB_{e}}\,$. In addition, in the amplitude ${\cal
M}_{\pi^-\to\, l\,\bar\nu_l}$ one should take $q^{3}=-k^{3}$.

\subsection{Evaluation of the weak decay amplitude}

As a second step we concentrate on the evaluation of the sum of the squared
amplitudes for all possible lepton polarizations, $\overline{\big|{\cal
M}_{\pi^-\to\, l\,\bar\nu_l}\big|^2}\,$. Using the properties of electron
and neutrino spinors in Eqs.~(\ref{sumspinnu}) and (\ref{sumspinl}), a
somewhat tedious but straightforward calculation leads to
\begin{eqnarray}
& & \overline{\big|{\cal
M}_{\pi^-\to\, l\,\bar\nu_l}\big|^2} \ = \  2\, G_F^2 \cos^2 \theta_c\,  \nonumber \\
& & \times\;\bigg\{ (E_l-q^3)\Big[ (E_{\bar\nu_l}-k^3)\,|M_{\parallel, +}^{-,+}|^2\,+\,
(E_{\bar\nu_l} + k^3)\,|M_{\perp,+}^{-,-}|^2\,-
\,2\,{\rm Re}\big( (M_{\parallel,+}^{-,+})^\ast M_{\perp,+}^{-,-}\, k^+\big)\Big] \nonumber \\
& & +\, (E_l+q^3)\Big[ (E_{\bar\nu_l}+k^3)\,|M_{\parallel, -}^{-,-}|^2\,+\,
(E_{\bar\nu_l} - k^3)\,|M_{\perp,-}^{-,+}|^2\,-
\,2\,{\rm Re}\big( (M_{\parallel,-}^{-,-})^\ast M_{\perp,-}^{-,+}\, k^- \big)\Big] \nonumber \\
& & +\,2\,\sqrt{2nB_e}\, \Big[
(E_{\bar\nu_l} + k^3)\,{\rm Im}\big((M_{\parallel,-}^{-,-})^\ast M_{\perp,+}^{-,-}\big)
- (E_{\bar\nu_l} - k^3)\,{\rm Im}\big((M_{\parallel,+}^{-,+})^\ast
M_{\perp,-}^{-,+}\big)
\nonumber \\
& & - \,{\rm Im}\big(M_{\parallel,+}^{-,+} (M_{\parallel, -}^{-,-})^\ast\, k^-\big)
\, - \,{\rm Im}\big(M_{\perp,+}^{-,-}
(M_{\perp,-}^{-,+})^\ast\, k^+\big)\Big]\bigg\}\ ,
\label{long}
\end{eqnarray}
where we have used the definitions
\begin{equation}
M_{\parallel,\lambda}^{-,\,\epsilon} \ = \ M_{\lambda}^{-,0} +\epsilon\, M_{\lambda}^{-,3}\ ,
\qquad
M_{\perp,\lambda}^{-,\,\epsilon} \ = \ M_{\lambda}^{-,1} + i\epsilon\, M_{\lambda}^{-,2} \ ,
\qquad
k^\epsilon \ = \ k^1 + i\epsilon\, k^2 \ .
\label{emesa}
\end{equation}

To proceed we need to calculate $M_{\parallel,\lambda}^{-,\,\epsilon}$ and
$M_{\perp,\lambda}^{-,\,\epsilon}$, which are related to the matrix elements of the
hadronic current through Eq.~(\ref{integral}). Defining
\begin{eqnarray}
& & H_{\parallel,L}^{-,\,\epsilon} \ = \ H_{L}^{-,0} + \epsilon\, H_{L}^{-,3} \
= \ H_{\parallel,V}^{-,\,\epsilon} - H_{\parallel,A}^{-,\,\epsilon}\ , \nonumber \\
& & H_{\perp,L}^{-,\,\epsilon} \ = \ H_{L}^{-,1} + i\epsilon\, H_{L}^{-,2}\ =
\ H_{\perp,V}^{-,\,\epsilon} - H_{\perp,A}^{-,\,\epsilon}\ ,
\label{hls}
\end{eqnarray}
from the expressions in Eqs.~(\ref{hvpar}-\ref{hav}) we get
\begin{eqnarray}
H^{-,\,\epsilon}_{\parallel,L} (x,\bar p) & = & i\,\sqrt{2}\, (f^{(A1)}_{\pi^-}
-\epsilon\, f^{(V)}_{\pi^-})  (E_{\pi^-} +\epsilon\, \,p^3)
\,\mathbb{F}^-_{\bar p}(x)\ ,
\nonumber\\
H^{-,\,\epsilon}_{\perp,L} (x,\bar p) & = & -\epsilon\,\sqrt{2}\,
\big(f^{(A1)}_{\pi^-} +\epsilon\, f^{(A2)}_{\pi^-} - f^{(A3)}_{\pi^-} \big)
\sqrt{B_e\left( 2\ell+1+\epsilon \right)}\;
\mathbb{F}^-_{\bar p+\epsilon}(x)\ .
\label{hepsilon}
\end{eqnarray}
Now the integral over space-time variables in Eq.~(\ref{integral}) can be
carried out as described in Appendix C. In this way one gets
\begin{eqnarray}
\hspace{-0.5cm} M_{\parallel,\lambda}^{-,\,\epsilon} & = &
i\,\sqrt{2}\, (f^{(A1)}_{\pi^-} -\epsilon\, f^{(V)}_{\pi^-})  (E_{\pi^-} +\epsilon \,p^3)\,
{\cal G}^{\,-,\lambda}_{\ell, n}(k_\perp)\,e^{ik^1(p^2+q^2)/(2B_e)}\ ,
\nonumber \\
\hspace{-0.5cm} M_{\perp,\lambda}^{-,\,\epsilon} & = & -\epsilon\,\sqrt{2}\,
\big(f^{(A1)}_{\pi^-} +\epsilon\, f^{(A2)}_{\pi^-} - f^{(A3)}_{\pi^-} \big)
\sqrt{B_e(2\ell+1+\epsilon)} \
{\cal G}^{\,-,\lambda}_{\ell+\epsilon,n}(k_\perp)\,e^{ik^1(p^2+q^2)/(2B_e)} \ ,
\label{emes}
\end{eqnarray}
where the functions $\mathcal{G}^{\,-,\lambda}_{\ell,n}(k_\perp)$ are
explicitly given in Eq.~(\ref{Mcal}). Notice that
$M_{\parallel,\lambda}^{-,\,\epsilon}$ and $M_{\perp,\lambda}^{-,\,\epsilon}$
depend on $p^2$ and $q^2$ only through a global common phase, therefore the
decay width turns out to be $p^2$ and $q^2$ independent.

Let us consider once again the case in which the decaying pion is in the
state of lowest energy. Taking $\ell=0$, from Eq.~(\ref{Mcal}) one gets
\begin{eqnarray}
{\cal G}^{\,-,\pm}_{-1,n}(k_\perp) & = & 0 \ ,\nonumber \\
{\cal G}^{\,-,-}_{0,n}(k_\perp) & = & 2\pi\, \exp\Big(-\frac{k_\perp^2}{4B_e}\Big)
\frac{1}{\sqrt{n!}}\left(\frac{-ik^-}{\sqrt{2B_e}}\right)^n \ , \nonumber \\
{\cal G}^{\,-,+}_{0,n}(k_\perp) & = & (1-\delta_{n0})\, {\cal G}^{\,-,-}_{0,n-1}(k_\perp) \ , \nonumber \\
{\cal G}^{\,-,-}_{1,n}(k_\perp) & = & 2\pi\, \exp\Big(-\frac{k_\perp^2}{4B_e}\Big)
\frac{1}{\sqrt{n!}}\left(\frac{-ik^-}{\sqrt{2B_e}}\right)^{n-1} \left(n-\dfrac{k_\perp^2}{2B_e}\right)
\ , \nonumber \\
{\cal G}^{\,-,+}_{1,n}(k_\perp) & = & (1-\delta_{n0})\, {\cal G}^{\,-,-}_{1,n-1}(k_\perp)\ .
\label{mmasmen}
\end{eqnarray}
The result obtained for the sum of the squared amplitudes, taking $\ell
= 0$ and $p^3=0$, can be replaced in the expression for the partial decay
width $\Gamma_l^-(B)$, Eq.~(\ref{gamgen-1}).
Notice that for a pion in the lowest energy state the ``perpendicular''
amplitudes $M_{\perp,\lambda}^{-,\,\epsilon}$ vanish for $\epsilon = -1$.
Thus, from Eq.~(\ref{emes}), it is seen that the partial decay width will
depend on three form factor combinations, namely
\begin{equation}
a_{\pi^-} = f^{(A1)}_{\pi^-} - f^{(V)}_{\pi^-}\ , \qquad b_{\pi^-} = f^{(A1)}_{\pi^-} + f^{(V)}_{\pi^-}\ ,
\qquad c_{\pi^-} = f^{(A1)}_{\pi^-} + f^{(A2)}_{\pi^-} - f^{(A3)}_{\pi^-}\ .
\label{defabc}
\end{equation}
We recall that the form factors are in general functions of $\ell$, $p^3$
and $B$ (to be evaluated in this case at $\ell = 0$, $p^3=0$). In this way,
after some algebra one arrives at
\begin{eqnarray}
\Gamma_l^-(B) & = &
\frac{G_F^2\cos^2\theta_c}{2\pi\,E_{\pi^-}^2}\, B_e\,\sum_{n=0}^{n_{\rm
max}}\int_0^{x_{\rm max}}dx\  \frac{1}{\rule{0cm}{0.4cm}\bar{k^3}}\;\frac{x^{n-1}}{n!}\
e^{-x} \;A^{(n)}_{\pi^-}(x)\ ,
\label{gamgenfinal}
\end{eqnarray}
where we have introduced the dimensionless variable $x=k_\perp^2/(2B_e)$ and
used the definitions
\begin{eqnarray}
A^{(n)}_{\pi^-}(x) & = & \big[E_{\pi^-}^2 -2B_e(n-x)-m_l^2\big]  \nonumber \\
&& \times \left[\frac{m_l^2}{2}\,(n |a_{\pi^-}|^2 + x |b_{\pi^-}|^2) + B_e (n-x)
(n|a_{\pi^-} - c_{\pi^-} |^2+x|b_{\pi^-} - c_{\pi^-}|^2)\right]
\nonumber \\
& & + 2B_ex
\big[E_{\pi^-}^2(n|a_{\pi^-} - b_{\pi^-} |^2-(n-x)|b_{\pi^-} - c_{\pi^-} |^2)+(n-x)\,m_l^2|c_{\pi^-} |^2\,\big]
\end{eqnarray}
and
\begin{equation}
x_{\rm max} \ = \ \frac{(E_{\pi^-}-\sqrt{2nB_e+m_l^2})^2}{2B_e}\ .
\end{equation}
In addition, from Eq.~(\ref{k3}) one has
\begin{equation}
\bar{k^3} \ = \
\frac{1}{2E_{\pi^-}}\left\{\Big[E_{\pi^-}^2-2B_e(n-x)-m_l^2\Big]^2
-8B_e\,E_{\pi^-}^2\,x\right\}^{1/2}\ ,
\end{equation}
while $n_{\rm max}$ is given by Eq.~(\ref{nmax}). Notice that for $\ell=0$,
$p^3=0$, the $\pi^-$ energy is given by $E_{\pi^-} = \sqrt{B_e +
m_{\pi^-}^2}$.

It is worth remarking that the above expression for $\Gamma_l^-(B)$
corresponds to the case $B>0$, $s=\sigma=-$, in Eqs.~(\ref{hvpar}-\ref{hav}).
For $B<0$, $s=-\sigma=+$, we obtain an expression for $\Gamma_l^-(B)$
similar to that in Eq.~(\ref{gamgenfinal}), where in the function
$A^{(n)}_{\pi^-}(x)$ one has to exchange $a_{\pi^-}\leftrightarrow b_{\pi^-}$.
Recalling that $f^{(V)}_{\pi^-}$ has to be an odd function of $B$, it is
seen that the partial decay width is invariant under the exchange $B\to -B$,
as expected.

\subsection{Large magnetic field limit}

It is interesting to study the case of a large external magnetic field. As
stated, since the pion is built with charged quarks, the pion mass will
depend in general on the magnetic field. Now, if the mass growth is
relatively mild, for large magnetic fields one should get $B_e >
m_\pi^2-m_l^2$. In fact, this is what is obtained from lattice QCD
calculations~\cite{Bali:2018sey} as well as from effective approaches like
the Nambu--Jona-Lasinio model~\cite{Coppola:2018vkw}, for values of $B_e$ say
$\agt 0.05$~GeV$^2$. According to Eq.~(\ref{nmax}), this implies $n_{\rm
max} = 0$; hence the outgoing muon or electron (assuming that the energy is
below the $\tau$ production threshold) has to lie in its lowest Landau level
$n=0$. In this case the expression for the partial width simplifies to
\begin{eqnarray}
\Gamma_{l,\,0}^{-}(B) & = &
\frac{G_{F}^{2}\cos^{2}\theta_{c}}{4\pi E_{\pi^-}} \! \int_0^{E_{\pi^-}-\,m_l}
dk_{\perp}\, k_{\perp}\;
\frac{e^{-k_{\perp}^{2}/(2B_{e})}}{\bar k^3}
\nonumber \\
& & \times\,
\bigg[ m_l^2\Big(\!E_{\bar\nu_l} |b_{\pi^-}|^2 -
\frac{k_\perp^2}{E_{\pi^-}}|c_{\pi^-}|^2\Big)
 + E_l\,k_\perp^2 | b_{\pi^-} - c_{\pi^-} |^2\bigg]\ ,
\label{unno}
\end{eqnarray}
where $E_{\bar\nu_l} = E_{\pi^-}-E_l = \sqrt{{\bar
k}^3{\rule{0cm}{.35cm}}^2+k_\perp^2}$ and $E_l = \sqrt{ {\bar
k}^3{\rule{0cm}{.35cm}}^2 + m_l^2}$, with
\begin{equation}
\bar{k^3} \ = \
\frac{1}{2E_{\pi^-}}\left[\left(E_{\pi^-}^2+k_{\perp}^{2} -m_l^2\right)^2-4 \,E_{\pi^-}^2\,k_{\perp}^2\right]^{1/2}\ .
\end{equation}

A further simplification can be obtained in the case where the squared
lepton mass can be neglected in comparison with $B_{e}$ (or, equivalently,
in comparison with $E_{\pi^-}^2$, which is expected to grow approximately
as $B_{e}$). Setting $m_l=0$, one has
$E_l = \bar k^3$, and the integral over $k_\perp$ extends up to
$E_{\pi^-}$. Thus, the decay width is given by
\begin{eqnarray}
\Gamma_{l,\,0}^{-}(B)\Big|_{m_l=0} & = &
\frac{G_{F}^{2}\cos^{2}\theta_{c}}{2\pi
E_{\pi^-}}\int_0^{E_{\pi^-}}\!\!\! dk_{\perp} \, k_{\perp}^{3} \,
e^{-k_{\perp}^{2}/(2B_{e})}\,|\,b_{\pi^-} - c_{\pi^-}|^{2} \nonumber
\\
 &=&\
\frac{G_{F}^{2}\cos^{2}\theta_{c}}{\pi}\;\frac{B_{e}^{2}}{E_{\pi^-}}
\Big[1-\Big(1+\frac{E_{\pi^-}^{2}}{2B_{e}}\Big)\,
e^{-E_{\pi^-}^{2}/(2B_{e})}\,\Big] \,|b_{\pi^-} - c_{\pi^-}|^{2} \ , \label{60}
\end{eqnarray}
where, according to the definitions in Eq.~(\ref{defabc}),
\begin{equation}
b_{\pi^-} - c_{\pi^-} \ = \ f^{(V)}_{\pi^-}- f_{\pi^-}^{(A2)}+
f_{\pi^-}^{(A3)}\ .
\label{b-c}
\end{equation}
It is worth stressing that the decay width does not vanish in the limit
$m_l=0$; i.e.~it does not show the helicity suppression found in the $B=0$
case. In fact, it turns out to grow with the magnetic field as
$B_e^2/E_{\pi^-}$, with some suppression due to the factor in square
brackets. Moreover, it is seen that the contributions of the usual pion
decay form factor, $f_{\pi^-}^{(A1)}$, get canceled in Eq.~(\ref{b-c});
hence the decay width is proportional to a combination of form factors that
do not contribute to the hadronic amplitude in the absence of the external
magnetic field. Clearly, the relevance of Eq.~(\ref{60}) depends on whether
these form factors are non-negligible for magnetic fields that are much
larger than the lepton mass squared. While this is likely to happen for the
$\pi^-$ decay to $e\,\bar \nu_e$, in the case of the muon (and of course,
the tau) the situation is less clear, and the corrections arising from a
finite lepton mass should be taken into account. Interestingly, it is
possible to obtain relatively simple expressions for the $\pi^-\to l\bar
\nu_l$ decay width at leading order in the ratio $m_l/E_{\pi^-}$. From
Eq.~(\ref{unno}) one gets
\begin{eqnarray}
\Gamma_{l,\,0}^{-}(B) & = & \Gamma_{l,\,0}^{-}(B)\Big|_{m_l=0} +
\frac{G_{F}^{2}\cos^{2}\theta_{c}}{2\pi}\, \frac{B_e\, e^{-E_{\pi^-}^{2}/(2B_{e})}}{E_{\pi^-}}
\nonumber \\
& & \times \Big[ f_1\,
|b_{\pi^-}|^2 - 2 f_2 \, {\rm Re}(b_{\pi^-}^\ast \ c_{\pi^-}) +
f_3\, |c_{\pi^-}|^2 \Big]\, m_l^2\  + \ {\cal O}\Big(\frac{m_l^3}{E^3_{\pi^-}}\Big)\
,
\label{ttres}
\end{eqnarray}
where
\begin{eqnarray}
f_1 &=& (1+\alpha)^2  - (1 + 2\alpha)\,e^{\alpha} + 2 \alpha^2
\left(I(\alpha) - \ln \frac{m_l}{E_{\pi^-}}\right) \ , \\
f_2 &=& \alpha(2+\alpha) - 2\alpha\, e^{\alpha}
+ 2\alpha (\alpha-1) \left(I(\alpha) - \ln \frac{m_l}{E_{\pi^-}}\right) \ ,\\
f_3 &=& \alpha^2+2\alpha-2 + 2 (1-\alpha)\, e^{\alpha}+ 2\alpha (\alpha-2)
\left(I(\alpha) - \ln \frac{m_l}{E_{\pi^-}}\right) \ ,
\end{eqnarray}
with $\alpha = E_{\pi^-}^2/2 B_e$ and $I(\alpha) = \int_0^1 dx (e^{\alpha x}
- 1)/x$. It is seen that for $m_l = m_\mu = 105.65$~MeV and $B_e \agt
0.3$~GeV$^2$, Eq.~(\ref{ttres}) approximates the full result in
Eq.~(\ref{unno}) within 15\% accuracy.

It is also interesting at this point to compare our result in
Eq.~(\ref{unno}) with the expression quoted in Eq.~(5) of
Ref.~\cite{Bali:2018sey}, which also corresponds to the limit of a large
external magnetic field. The authors of that work make some approximations
for the motion of a charged pion in the presence of the magnetic field,
concluding that only one of the two possible antineutrino polarizations can
contribute to the decay amplitude. Moreover, based on considerations of
angular momentum conservation, they assume that the antineutrino
momentum in the perpendicular plane $\vec k_\perp$ vanishes. In fact,
it can be seen that if one imposes such a condition in Eq.~(\ref{unno}), the
result quoted in Ref.~\cite{Bali:2018sey} can be recovered. However, we find
that if one takes full account of the effect of the magnetic field on
charged pion wave functions, conservation laws do not imply $\vec
k_\perp =0$; therefore one should integrate over all possible values of the
antineutrino momentum, as in Eq.~(\ref{unno}). Another main difference
between our work and the analysis in Ref.~\cite{Bali:2018sey} is that our
calculations include a perpendicular piece of the hadronic amplitude
(related to $c_{\pi^-}$), which arises due to the presence of a $\pi^-$ zero
point motion in the perpendicular plane, even in the $\ell=0$ state.

The study of angular momentum in the presence of magnetic fields has
been addressed in the recent literature (see
e.g.~Ref.~\cite{Wakamatsu:2017isl}) and deserves some discussion. As
suggested in Ref.~\cite{Wakamatsu:2017isl}, the consequences of the axial
symmetry of the problem, as well as the physical meaning of angular momenta,
can be better understood if one works in the symmetric gauge. Having this in
mind, we have rederived the $\pi^-\to l\bar\nu_l$ decay width in this gauge,
considering for simplicity the case of a charged pion in its lowest energy
state and a charged lepton in the lowest Landau level. For the spatial wave functions of the
pion and charged lepton states we have used the functions given
e.g.~in Eq.~(17) of Ref.~\cite{Wakamatsu:2017isl} (see also Ref.~\cite{Sokolov:1986nk}).
These functions are eigenstates of the $z$ component of the orbital angular momentum operator,
$L^3$, with eigenvalues say $M_p$ and $M_q$ for the pion and the charged
lepton, respectively. If the particles are in the lowest Landau level, it is found that
$M_p$ and $M_q$ can take any integer value~$\leq 0$. In order to deal with
right-handed antineutrino states we have used spinors that are eigenstates of
the $z$ component of the total angular momentum, $J^3$, with eigenvalue
$j^3$. Replacing in the corresponding weak decay matrix element and
performing the integral over space-time coordinates we obtain
\begin{eqnarray}
\langle l(q^3,n=0,M_q)\, \bar\nu_l(k^3,k_\perp,j^3) | {\cal L}_W |
\pi^-(p^3,\ell=0,M_p) \rangle & = & \hspace{5.8cm}\nonumber \\
& & \hspace{-5cm} (2\pi)^3
\, \delta(E_{\pi^-}-E_l-E_{\bar\nu_l})\, \delta(p^3-q^3-k^3) \, \delta_{M_p, M_q-\frac{1}{2} + j^3} \, {\cal M}\ ,
\end{eqnarray}
where ${\cal M}$ is a function that depends on the pion decay form factors
$f_{\pi^-}^{(V,A_i)}$ and on particle quantum numbers. Hence, in this gauge
it is clearly seen that angular momentum conservation leads to the selection
rule $M_p = M_q-1/2+j^3$, without requiring $k_\perp = |\vec k_\perp| = 0$.
Using the explicit form of ${\cal M}$ and taking
$p^3=0$, we have then calculated the sum/integral over all allowed outgoing
states. The outcome (which, as expected, does not depend on $M_p$), leads to
an expression for the decay width that coincides exactly with the one quoted in
Eq.~(\ref{unno}), confirming the gauge independence of our result. It
is also worth mentioning that, in addition to the usual ``canonical''
angular momentum, one can define ``mechanical'' angular momenta replacing
particle momenta $P^\mu$ by $P_{\rm mech}^\mu \equiv P^\mu-Q{\cal A}^\mu$
(a detailed comparison of both quantities for various gauge choices is given
in Ref.~\cite{Wakamatsu:2017isl}). Interestingly, for the $\pi^-\to
l\bar\nu_l$ decay in the limit of large $B$ (when only the $n=0$ state
contributes) it is found that the third components of mechanical angular
momenta of incoming and outgoing states only coincide when the transverse
antineutrino momentum vanishes, i.e., for $|k_\perp| = 0$.

\subsection{$B\rightarrow0$ limit}

In the $B\rightarrow 0$ limit our expressions should reduce to the
well-known result for the $\pi^- \to l\,\bar\nu_l$ decay width obtained in the absence
of external fields. For simplicity we consider the decay of a $\pi^{-}$ in
its fundamental state, taking $\ell=0$ and $p^{2}=p^{3}=0$. Then, in the $B\to 0$
limit the decay width should reduce to that of a $\pi^{-}$ at rest, namely
\begin{eqnarray}
\Gamma_l^-(0) & = & \frac{1}{2m_{\pi}}\int\frac{d^{3}k}{\left(2\pi\right)^{3}2E_{\bar\nu_l}}\,
\frac{d^{3}q}{\left(2\pi\right)^{3}2E_{l}}\,\left(2\pi\right)^{4}\delta^{(4)}
(p-q-k)\nonumber \\
& & \times\, \big(G_F\,\cos\theta_c\,f_{\pi}\big)^{2}\; 8\,m_{\pi}^{2}\,
\big(E_{l}\,E_{\bar\nu_l}+\vec{q}\cdot\vec{k}\big)
\label{V.1} \\
& = & \rule{0cm}{0.9cm}\frac{\big(G_F\,\cos\theta_c\,f_{\pi}\big)^2}{4\pi}\; m_\pi\,m_l^2\,
\bigg(1-\frac{m_l^2}{m_\pi^2}\bigg)^2\ ,
\label{bzero}
\end{eqnarray}
where $p^{\mu}=(m_{\pi},\vec{0})$; $q^{\mu}=(E_{l},\vec{q})$; and
$k^{\mu}=(E_{\bar\nu_l},\vec{k})$ stand for the pion, lepton and
antineutrino four-momenta, respectively. It can be checked that for a given
value of the lepton mass the numerical results for Eqs.~(\ref{bzero}) and
(\ref{gamgenfinal}), in the limit $B_e\to 0$, $n_{\rm max}\to\infty$,
$a_{\pi^-} = b_{\pi^-} = c_{\pi^-} =f_\pi$, are coincident. However, the
comparison between Eqs.~(\ref{gamgen}) and (\ref{V.1}) still leads to the
question of how one can arrive at four-momentum conservation in the limit of
vanishing external magnetic field. Hence, the goal of this section is to
recover analytically the Dirac deltas of momentum conservation, obtaining
Eq.~(\ref{V.1}) from Eq.~(\ref{gamgen}) in the $B\rightarrow0$ limit.

The presence of a magnetic field implies the existence of a characteristic
time and a characteristic length of the system, given in natural units by
$B_{e}^{-1/2}$, which is usually called the ``magnetic length''. In
the $B\to 0$ limit these have to be much larger than the time $T$ and the
distance $2L$ along which the interaction is active, i.e.
\begin{equation}
\sqrt{B_{e}}\,L\ \ll\ 1  \ ,\qquad\qquad \sqrt{B_{e}}\,T\ \ll\ 1 \ .
\label{V.2}
\end{equation}
In addition, we assume that $T$ and $L$ are large enough so that the product
of any momentum of the system times $L$ or any energy of the system times
$T$ turns out to be much larger than~1. In particular,
\begin{equation}
\sqrt{nB_{e}}\,L\gg 1\ , \qquad k_{x}\,L\gg 1 \ ,\qquad
E_{l}\,T \gg 1\ , \qquad E_{\bar\nu}\,T\gg 1\ .
\label{V.3}
\end{equation}
We recall that the lepton energy is given by
$E_{l}=\sqrt{m_{l}^{2}+2nB_{e}+(q^{3})^{2}}$, which implies that the
magnetic field will contribute significantly to the energy only for very
large values of $n$. In fact, from this expression we can infer that the
term $2nB_{e}$ will lead to the $\left(q_{\perp}\right)^{2}$ contribution to
the energy in the $B\rightarrow0$ limit. As $B$ decreases, the contributing
leptonic states will have increasingly larger values of $n$, in such a way
that $nB$ remains finite.

From Eqs.~(\ref{vecop}), (\ref{hvch}) and (\ref{hva}), we observe that in
the $B\rightarrow0$ limit the only nonvanishing pion decay form factor is
$f_{\pi^-}^{(A1)}$. Thus, from Eqs.~(\ref{long}-\ref{hepsilon}) we have
\begin{eqnarray}
\hspace{-0.7cm}\overline{\big|{\cal M}_{\pi^-\to\, l\,\bar\nu_l}\big|^2}\;\Big|_{\ell=p^3=0}
 & = & 4\, G_F^2 \cos^2 \theta_c\, \left|f_{\pi^-}^{(A1)}\right|^{2} \nonumber \\
 &  & \hspace{-3.7cm} \times\;\bigg\{E_{\pi^-}^2\, (E_{l}-q^{3})\,(E_{\bar\nu_{l}}-k^{3})\,|\,{\cal I}_{0,n}^{\,-,+}|^{2}
 +E_{\pi^-}^2\,(E_{l}+q^{3})\,(E_{\bar\nu_{l}}+k^{3})\,|{\cal I}_{0,n}^{\,-,-}|^{2}\nonumber \\
 &  & \hspace{-3.7cm}+\; 2B_{e}\,(E_{l}+q^{3})\,(E_{\bar\nu_{l}}-k^{3})\,|{\cal I}_{1,n}^{\,-,-}|^{2}\,
 -\,2\,\sqrt{2B_{e}}\,E_{\pi^-}\,(E_{l}+q^{3})\,{\rm Im}\big(({\cal I}_{0,n}^{\,-,-})^{\ast}
 \,{\cal I}_{1,n}^{\,-,-}\,k^-\big)+\nonumber \\
 &  & \hspace{-3.7cm} + \,2\,\sqrt{2nB_e}\,E_{\pi^-}\,\Big[\sqrt{2B_{e}}\,(E_{\bar\nu_{l}}-k^{3})
 \,{\rm Im}\big(({\cal I}_{0,n}^{\,-,+})^{\ast}\,{\cal I}_{1,n}^{\,-,-}\big)-
 E_{\pi^-}\,{\rm Im}\big({\cal I}_{0,n}^{\,-,+}\,({\cal
 I}_{0,n}^{\,-,-})^\ast\,k^-\big)\,\Big]\bigg\}\ ,
\label{widthi}
\end{eqnarray}
where the functions ${\cal I}_{\ell,n}^{\,-,\pm}$ are given by
\begin{eqnarray}
\mathcal{I}_{\ell,n}^{\,-,\pm} & = &
N_{\ell}\,N_{n_\mp}\int_{-L}^{L}dx^{1}\,e^{ik^{1}x^{1}}D_{\ell}\left(\sqrt{2B_{e}}\,x^{1}\right)
D_{n_\mp}\left(\sqrt{2B_{e}}\,x^{1}+\sqrt{\dfrac{2}{B_{e}}}\,q^{2}\right)\ .
\label{V.5}
\end{eqnarray}
In fact, the latter correspond to the functions ${\cal
I}^{\,-,\lambda}_{\ell, n}(k_\perp,p^2,q^2)$ defined in Eq.~(\ref{inti}),
taking $p^2=0$ and restricting the integral over $x^1$ to the interval
$(-L,L)$ in order to take into account the conditions in Eqs.~(\ref{V.2})
and (\ref{V.3}).

Owing to the restriction of the integration interval in Eq.~(\ref{V.5}), in
the $B\to 0$ limit one has $\sqrt{2B_{e}}\,x^{1}\to 0$. Thus, the factors
$N_{\ell}\,D_{\ell}(\sqrt{2B_{e}}\,x^{1})$, where $\ell=0$ or 1, satisfy
\begin{eqnarray}
N_{0}D_{0}\left(\sqrt{2B_{e}}x^{1}\right) &
\underset{{\scriptscriptstyle \sqrt{2B_{e}}x^{1}\rightarrow0}}{\longrightarrow} &
\left(4\pi B_{e}\right)^{1/4}\ \ ,
\label{dlim}\\
N_{1}D_{1}\left(\sqrt{2B_{e}}x^{1}\right) &
 \underset{{\scriptscriptstyle \sqrt{2B_{e}}x^{1}\rightarrow0}}{\longrightarrow}
 & \left(4\pi B_{e}\right)^{1/4}\frac{1}{2}\sqrt{2B_{e}}x^{1}\ \sim \ 0\ \ ,
\end{eqnarray}
and the terms with ${\cal I}_{1,n}^{\,-,-}$ in Eq.~(\ref{widthi}) can be
neglected. We obtain
\begin{eqnarray}
\overline{\big|{\cal M}_{\pi^-\to\, l\,\bar\nu_l}\big|^2}\;\Big|_{\ell=p^3=0}
 & = & 4\, G_F^2 \cos^2 \theta_c\, \left|f_{\pi^-}^{(A1)}\right|^{2}
 E_{\pi^-}^{2}\, \bigg\{(E_{l}-q^{3})\,(E_{\bar\nu_{l}}-k^{3})\,\left|{\cal I}_{0,n}^{\,-,+}\right|^{2} \nonumber \\
 &  & \hspace{-2.5cm}+\; (E_{l}+q^{3})\,(E_{\bar\nu_{l}}+k^{3})\,\left|{\cal I}_{0,n}^{\,-,-}\right|^{2}
 -2\,\sqrt{2nB_e}\,\,{\rm Im}\big({\cal I}_{0,n}^{\,-,+}({\cal I}_{0,n}^{\,-,-})^\ast\,k^-\big)\bigg\}\
 .
\label{V.11}
\end{eqnarray}
The detailed calculation of the functions ${\cal I}_{0,n}^{\,-,\lambda}$
and their contributions to the expression in Eq.~(\ref{V.11}) is given in
Appendix D. We finally arrive at
\begin{eqnarray}
\overline{\big|{\cal M}_{\pi^-\to\, l\,\bar\nu_l}\big|^2}\;\Big|_{\ell=p^3=0}
 & = & 8\, G_F^2 \cos^2 \theta_c\, \left|f_{\pi^-}^{(A1)}\right|^{2} \,
 4\pi L\left(4\pi B_{e}\right)^{1/2}\,
\theta\left(1-\frac{\left|q^{2}\right|}{\sqrt{2nB_{e}}}\right)
 \frac{B_{e}}{\bar{q}_{n}} \nonumber \\
 & & \times \,
 E_{\pi^-}^{2}\;\Big\{\left(E_{l}E_{\bar\nu_{l}}+q^{2}\,k^{2}+q^{3}\,k^{3}\right)\,\left[
 \delta (k^{1}-\bar{q}_{n})+\delta(k^{1}+\bar{q}_{n})\right]\nonumber \\
 &  & +k^{1}\,\bar{q}_{n}\,\delta (k^{1}+\bar{q}_{n})-
 k^{1}\,\bar{q}_{n}\, \delta(k^{1}-\bar{q}_{n})\Big\}\ ,
\end{eqnarray}
where $\bar{q}_{n} \ = \ \sqrt{2nB_{e}-\left(q^{2}\right)^{2}}$. We recall
that $n$ has to be a large number, in such a way that $nB_e$ is kept finite
for small $B$.

We introduce now a new variable $q^{1}$, in the following way:
\begin{eqnarray}
\overline{\big|{\cal M}_{\pi^-\to\, l\,\bar\nu_l}\big|^2}\;\Big|_{\ell=p^3=0}
 & = & 8\, G_F^2 \cos^2 \theta_c\, \left|f_{\pi^-}^{(A1)}\right|^{2} \,
 2L\left(4\pi B_{e}\right)^{1/2}\,
 \theta\left(1-\frac{\left|q^{2}\right|}{\sqrt{2nB_{e}}}\right)
 \frac{B_{e}}{\bar{q}_{n}} \nonumber \\
 & & \times \int\, dq^1\, 2\pi\delta (q^1+k^1) \,
\left[\delta(q^{1}-\bar{q}_{n})+\delta(q^{1}+\bar{q}_{n})\right] \nonumber \\
 & & \times \; E_{\pi^-}^{2}\,\left(E_{l}E_{\bar\nu_{l}}+q^{1}\,k^{1}+q^{2}\,k^{2}+q^{3}\,k^{3}\right)
\ .
\end{eqnarray}
Next, let us consider the decay width in Eq.~(\ref{gamgen}). We need to
treat with some care the pion density $n_{\pi^-}$, which appears in the
definition of the width in Eq.~(\ref{decB}). In fact, for a finite space
length $2L$, taking into account the approximation in Eq.~(\ref{dlim}), the
pion density will be given by
\begin{equation}
n_{\pi^-}=S\int_{-L}^{L}dx^{1}\,2\,E_{\pi^-}\,\left(4\pi B_{e}\right)^{1/2}=4\,L\,S\,E_{\pi^-}\left(4\pi
B_{e}\right)^{1/2}\ ,
\label{piondens}
\end{equation}
This result can be understood by writing the pion density in the form
$n_{\pi^-}=V\,2E_{\pi^-}\left(4\pi B_{e}\right)^{1/2},$ where $V$ is the
volume in which the interaction occurs. It is seen that Eq.~(\ref{piondens})
recovers the pion density in the absence of the magnetic field [see
Eq.~(\ref{pi0dens})], times a factor $(4\pi B_{e})^{1/2}$. The latter
compensates the fact that in the limit of small $B$, according to our
normalization of charged pion states, the spacial wave function of a pion in
a zero three-momentum state is
$N_{0}D_{0}\left(\sqrt{2B_{e}}x^{1}\right)\rightarrow\left(4\pi
B_{e}\right)^{1/4}$, instead of 1. Now, comparing the result in
Eq.~(\ref{piondens}) with the expression $n_{\pi^-}=4\pi S E_{\pi^-}$ quoted
in Eq.~(\ref{numberb}), it comes out that the width in the rhs of
Eq.~(\ref{gamgen}) has to be modified by a factor $\pi/[L\left(4\pi
B_{e}\right)^{1/2}]$. In addition, in the limit of small $B$ one can change
the sum over $n$ in Eq.~(\ref{gamgen}) by an integral over a variable
$\varkappa\equiv 2nB_{e}$. Hence, for $p^2=p^3=0$ we get
\begin{eqnarray}
\Gamma_{l}^{-}(B\to 0) & = & \frac{2\, G_{F}^{2}\cos^{2}\theta_{c}}{\pi E_{\pi^-}}\,\frac{1}{2B_{e}}
\int d\varkappa\int\frac{dq^{2}dq^{3}}{(2\pi)^{3}2E_{l}}\int\frac{d^{3}k}{(2\pi)^{3}2E_{\bar\nu_{l}}}
\,(2\pi)^{3}\,\nonumber \\
 &  & \times \; \delta(E_{\pi^-}-E_{l}-E_{\bar\nu_{l}})\,\prod_{i=2,3}\delta(q^{i}+k^{i})\,
 \frac{\pi}{L\left(4\pi B_{e}\right)^{1/2}}\nonumber \\
 &  & \times \, \int
 dq^{1}\left|f_{\pi^-}^{(A1)}\right|^{2}\,2\pi\delta(q^{1}+k^{1})\,2L
 \left(4\pi B_{e}\right)^{1/2}\,\theta\left(1-\frac{\left|q^{2}\right|}{\sqrt{\varkappa}}\right)
 \frac{B_{e}}{\sqrt{\varkappa-\left(q^{2}\right)^{2}}}\nonumber \\
  &  & \times \left[\delta\left(q^{1}-\sqrt{\varkappa - (q^2)^2}\right)+\delta\left(q^{1}+\sqrt{
 \varkappa - (q^2)^2}\right)\right]\nonumber \\
 &  & \times\;
 E_{\pi^-}^{2}\,\left(E_{l}E_{\bar\nu_{l}}+q^{1}\,k^{1}+q^{2}\,k^{2}+q^{3}\,k^{3} \right)
\ .
\end{eqnarray}
Finally, we can perform the integral over $\varkappa\,$. The delta functions
fix $q^{1}=\pm\sqrt{\varkappa-(q^{2})^{2}}$, leading to the expected result
$\varkappa=2nB_e=q_{\perp}^{2}$. Identifying $f_{\pi^-}^{(A1)}$ with $f_\pi$
in the $B\rightarrow 0$ limit, we arrive at the expression in
Eq.~(\ref{V.1}).

\section{Conclusions}

In this work we present a general method to parametrize the
one-pion-to-vacuum matrix elements of the vector and axial vector
hadronic currents in the presence of an external uniform static
magnetic field $\vec B$. Choosing this field to be orientated
along the 3-axis, we show that for the case of the neutral pion
the matrix elements of the parallel (0- and 3-) components of the
vector current can be expressed in terms of one single real form
factor, $f^{(V)}_{\pi^0}$, while the perpendicular (1- and 2-)
components vanish identically. For the matrix elements of the
axial vector current, two real form factors $f^{(A1)}_{\pi^0}$ and
$f^{(A3)}_{\pi^0}$ can be defined. Alternatively, the latter can
be written in terms of a parallel and a perpendicular form
factor, in consistency with the result obtained in
Ref.~\cite{Fayazbakhsh:2013cha}. In the case of the charged pion,
the situation is similar in what concerns the vector current.
Namely, the matrix elements of the parallel components can be
expressed in terms of a real (in general, nonvanishing) form
factor $f^{(V)}_{\pi^\sigma}$, common to both $\pi^+$ and $\pi^-$,
while the perpendicular (i.e.~1- and 2-) components vanish. This
is in agreement with the statement made in
Ref.~\cite{Bali:2018sey}. On the other hand, we find that three
form factors, $f^{(A1)}_{\pi^\pm}, f^{(A2)}_{\pi^\pm}$ and
$f^{(A3)}_{\pi^\pm}$, are in general required to parametrize the
matrix elements of the axial vector hadronic current. Once again,
the three of them are real, and they are equal for both pion
charges. The matrix elements of the charged pions in
Eqs.~(\ref{hvpar}-\ref{hav}) can be viewed as a proper
generalization of the corresponding expressions given in
Ref.~\cite{Bali:2018sey}. We have included here all possible gauge
covariant  structures, taking fully into account the effect of the
magnetic field on the charged pion wave functions. It should be
noticed that in the particular case of a charged pion lying in the
lowest Landau level, only two combinations of these three form
factors contribute to the decay width.

Using the above results we introduce a general, model-independent framework
to study the weak decay $\pi^- \to l\,\bar\nu_l$ in the presence of an
arbitrary large external magnetic field. For the case in which the decaying
pion lies in its state of minimum energy (i.e.~in the lowest Landau level,
with zero momentum along the 3-direction), we obtain an explicit expression
for the $\pi^- \to l\,\bar\nu_l$ decay width. The limits of this expression
for the cases of strong and weak magnetic fields are also studied, checking
that in the limit of $B=0$ it reduces to the usual result. It is interesting
to note that the expression obtained in the limit of large magnetic field,
Eq.~(\ref{unno}), is valid in most cases of physical interest. Namely, we
estimate its range of validity to be 0.05 GeV$^2 < eB < m_\tau^2 \approx 3$
GeV$^2$. It is seen that our result shows some differences with the one
given in Ref.~\cite{Bali:2018sey}, also obtained in the limit of large $B$.
We understand that the discrepancies arise from some approximations made in
Ref.~\cite{Bali:2018sey} concerning the motion of a charged pion in the
presence of the magnetic field. It is also worth noticing that the decay
width does not vanish in the limit $m_l = 0$; i.e.~it does not show the
helicity suppression found in the absence of the external magnetic field.

We finally mention that an obvious application of our work would be to study
how weak decay rates of charged pions get modified due to the presence of
the magnetic field. To reach this goal, however, the behavior of the decay
form factors as functions of the magnetic field should be determined. This
would require either the use of LQCD simulations, as proposed in
Ref.~\cite{Bali:2018sey}, or to rely on some hadronic effective model. We
expect to report soon on such a calculation in the framework of
Nambu--Jona-Lasinio-like models.

\section*{Acknowledgements}

This work has been supported in part by Consejo Nacional de Investigaciones Cient\'{i}ficas
y T\'{e}cnicas and Agencia Nacional de Promoci\'{o}n Cient\'{i}fica y Tecnol\'{o}gica (Argentina),
under Grants No. PIP14-492 and No. PICT14-03-0492; by the National University of La
Plata (Argentina), Project No. X824; by the Ministerio de Econom\'{i}a y Competitividad (Spain), under
Contract No. FPA2016-77177-C2-1-P; by the Centro de Excelencia Severo Ochoa
Programme, Grant No. SEV-2014-0398; and by Generalitat Valenciana (Spain), Grant No.
PrometeoII/2014/066.

\section*{Appendix A: Pion and fermion fields in a constant magnetic field}

\addtocounter{section}{1}
\setcounter{equation}{0}
\renewcommand{\theequation}{A\arabic{equation}}

In this appendix we quote expressions for the different fields used in our
work, written in terms of creation and annihilation operators. As in the
main text, we assume $\vec B = B \hat x^3$ and make use of the Landau gauge,
in which ${\cal A}^\mu = (0,0,B x^1, 0)$.

\subsection*{1. Neutral pion and neutrino fields}

The expressions for neutral fields do not get changed by the presence of the
external magnetic fields. Thus, they can be written in terms of the usual
creation and annihilation operators of definite momentum states. Following
the conventions given e.g.~in Ref.~\cite{Peskin}, the neutral pion field is
given by
\begin{equation}
\phi_{\pi^0}(x) \ = \   \int \frac{d^3p}{(2\pi)^3 2 E_{\pi^0}}
\left[ a(\vec p\,) \ e^{-i p\cdot x} + a^\dagger(\vec p\,) \ e^{i p\cdot x}
\right]\ ,
\label{bosonfield}
\end{equation}
where $x=(t, \vec x)$ and $p = (E_{\pi^0}, \vec p\,)$, with $E_{\pi^0}
= \sqrt{m_{\pi^0}^2 + |\vec p\,|^2}$ (it is worth mentioning that, in the
presence of an external field, one could also take into account corrections
leading to an anisotropic dispersion relation~\cite{Fayazbakhsh:2013cha}). 
The operators $a(\vec p\,)$ and $a^\dagger(\vec p\,)$ satisfy the commutation rule
\begin{equation}
[a(\vec p\,), a^\dagger(\vec p\,')] \ = \ 2 E_{\pi^0} \, (2\pi)^3 \, \delta^{(3)} (\vec p - \vec
p\,')\ .
\label{commbpart}
\end{equation}
For a finite volume $V$, the particle number $n_{\pi^0}$ associated with the momentum
eigenstate $|\pi^0(\vec p\,)\rangle = a^\dagger(\vec p\,) |0\rangle$ is given by
\begin{equation}
n_{\pi^0} \ = \ \int_V d^3x\; \langle \pi^0(\vec p\,) | \, j_{\pi^0}^{\,0}(t,\vec x) | \pi^0(\vec p\,)\rangle \
= \ 2 E_{\pi^0} V\ ,
\label{pi0dens}
\end{equation}
where
\begin{equation}
j_{\pi^0}^0(x)\ = \ i \left[ \phi_{\pi^0}^\dagger(x) \partial^0 \phi_{\pi^0}(x)
- \partial^0\phi_{\pi^0}^\dagger(x)\, \phi_{\pi^0}(x) \right]\ .
\label{pioncur}
\end{equation}

The neutrino field can be written as
\begin{equation}
\psi_{\nu_l}(x) \ = \   \sum_{r=1,2} \,
\int \! \frac{d^3k}{(2\pi)^3 2 E_{\nu_l}}
\left[ b(\vec k,r) \ U_{\nu_l}(x, \vec k,r) + d(\vec k,r)^\dagger  \ V_{\nu_l}(x,\vec k,r)
\right]\ ,
\label{neutrinofield}
\end{equation}
where
\begin{equation}
U_{\nu_l}(x,\vec k,r)  \ = \ u_{\nu_l}(\vec k,r) e^{-i k\cdot x}\ , \qquad \qquad
V_{\nu_l}(x,\vec k,r)  \ = \ v_{\nu_l}(\vec k,r) \ e^{i k\cdot x}\ .
\label{UVneut}
\end{equation}
Here $k = (E_{\nu_l},\vec k)$, with $E_{\nu_l}= |\vec k|$, while
$u_{\nu_l}(\vec k,r)$ and $v_{\nu_l}(\vec k,r)$ are the usual Dirac spinors with
polarization states $r=1$ or 2. They satisfy
\begin{equation}
\sum_{r=1,2} u_{\nu_l}(\vec k, r)\, \bar u_{\nu_l}(\vec k, r) \ = \ \sum_{r=1,2} v_{\nu_l}(\vec k, r)\, \bar v_{\nu_l}(\vec k, r)
 \ = \ \rlap/\!k\ .
\label{sumspinnu}
\end{equation}
Note that we are assuming that neutrinos are massless. The corresponding creation and
annihilation operators satisfy the relations
\begin{eqnarray}
\big\{b(\vec k,r), b(\vec k\,',r')^\dagger \big\} &=& \big\{d(\vec k,r), d(\vec k\, ',r')^\dagger \big\}
\: = \  2 E_{\nu_l}\, (2\pi)^3 \, \delta_{rr'}\, \delta^{(3)} (\vec k - \vec k\,')\ ,
\label{commf}
\\
\big\{b(\vec k,r), d(\vec k\,',r')\big\} &=& \big\{b(\vec k,r), d(\vec k\,',r')^\dagger \big\} \ = \ 0\ .
\label{comm}
\end{eqnarray}

\subsection*{2. Charged pion field}

The charged pion fields can be written as
\begin{equation}
\phi^s_{\pi^{\sigma}}(x) \ = \ \phi^s_{\pi^{-\sigma}}(x)^\dagger \ = \
\sum_{\ell=0}^\infty \int \frac{ dp^2  dp^3}{(2\pi)^3\, 2 E_{\pi^{\sigma}}}
\left[
a^\sigma(\breve p)  \; \mathbb{F}^{s}_{\bar p}(x) + a^{-\sigma}(\breve p)^\dagger  \; \mathbb{F}^{-s}_{\bar
p}(x)^\ast \right]\ ,
\label{chargepionexp}
\end{equation}
where  $Q_{\pi^\sigma}=\sigma |e|$ is the pion charge with $\sigma=\pm$, $s = \mbox{sign}(Q_{\pi^\sigma} B)$
and $B_e=|Q_{\pi^\sigma} B|$,. Note
that if $B>0$ then $s=\sigma$. As defined in the main text, $\bar p
= (p^0,\breve p)$, where $\breve{p} = (\ell,p^2,p^3)$ and the pion energy is given by $p^0= E_{\pi^{\sigma}} =
\sqrt{m^2_{\pi^{\sigma}}+ (2\ell+1) B_e + (p^3)^2}$.

The functions $\mathbb{F}^{s}_{\bar p}(x)$ are solutions of the eigenvalue
equation
\begin{equation}
{\cal D}_\mu {\cal D}^\mu \ \mathbb{F}^{s}_{\bar p} (x) \ = \ -  \left[ (p^0)^2 - (2 \ell+1) B_e - (p^3)^2 \right]
\mathbb{F}^{s}_{\bar p} (x)\ ,
\label{ecautovbo}
\end{equation}
where ${\cal D}^\mu = \partial^\mu + i s B_e \, x^1 \delta_{\mu 2}$. Their
explicit form is given by
\begin{equation}
\mathbb{F}_{\bar p}^s(x)\ = \
N_\ell\;e^{-i(p^0x^0-p^2x^2-p^3x^3)}\,D_\ell(\rho_{s})\ ,
\label{efes}
\end{equation}
where $N_\ell= (4\pi B_e)^{1/4}/\sqrt{\ell!}\,$, $\rho_{s} =
\sqrt{2B_e}\,x^1-s\sqrt{2/B_e }\,p^2$ and $D_\ell(x)$ are
cylindrical parabolic functions. The latter are defined as
\begin{equation}
D_\ell(x) \ = \ 2^{-\ell/2}\,e^{-x^2/4}\,H_\ell(x/\sqrt{2})\ ,
\label{dll}
\end{equation}
$H_\ell(x)$ being the Hermite polynomials, with the standard convention
$H_{-1}(x) = 0$. It can be seen that the functions $\mathbb{F}^{s}_{\bar
p}(x)$ satisfy the orthogonality relations
\begin{eqnarray}
\sum_{\ell=0}^\infty \int \frac{dp^0 dp^2  dp^3}{(2\pi)^4} \ \mathbb{F}_{\bar p}^{s}(x)\,\mathbb{F}^s_{\bar p}(x')^\ast
& = & \delta^{(4)}{(x- x')}\ ,
\\
\int d^4x \ \mathbb{F}_{\bar p'}^{s}(x)^\ast \  \mathbb{F}^s_{\bar p}(x) & = &
(2\pi)^4 \,\delta_{\ell\ell'} \, \delta(p^0-p^{\,\prime\, 0})\, \delta(p^2-p^{\,\prime\, 2}) \,\delta(p^3-p^{\,\prime\, 3})\  . \label{orthF}
\end{eqnarray}
In addition, the creation and annihilation operators in
Eq.~(\ref{chargepionexp}) satisfy the commutation relations
\begin{equation}
\left[a^\sigma(\breve p) , a^\sigma(\breve p\,')^\dagger\right]  \ = \ \left[a^{-\sigma}(\breve p) , a^{-\sigma}(\breve p\,')^\dagger\right] \ = \ 2 E_{\pi^{\sigma}}  \,(2\pi)^3  \, \delta_{\ell\ell'}\,
\delta(p^2-p^{\,\prime\, 2})\, \delta(p^3-p^{\,\prime\, 3})\ .
\label{conbos}
\end{equation}
Note that with these conventions the operators $a^\sigma(\breve p)$  and $a^{-\sigma}(\breve
p)$ turn out to have different dimensions from the creation and annihilation
operators that are usually defined in the absence of the external magnetic field
[and also from those corresponding to the $\pi^0$ field; see
Eq.~(\ref{commbpart})].

It is also useful to calculate the particle number associated with the state
$|\pi^\sigma(\breve p) \rangle = a^\dagger(\breve p) |0\rangle$ in a volume
$V$. Given our choice of the Landau gauge, in this case it is convenient to
consider an infinite cylinder of section $S$ lying along the $x^1$ axis.
We obtain
\begin{equation}
n_{\pi^\sigma} \ = \ \int_{-\infty}^\infty dx^1
\int_S dx^2\,dx^3\;
\langle \pi^\sigma(\breve p) |  \, j_{\pi^\sigma}^{\,0}(x) | \pi^\sigma(\breve p)\rangle
\ = \ 2 E_{\pi^{\sigma}} 2\pi S\ ,
\label{numberb}
\end{equation}
where the current is defined in a similar way to that corresponding to
the neutral pion; see Eq.~(\ref{pioncur}). Note that we are normalizing to
$4\pi E$ particles per unit surface, which differs from the usual
normalization $\rho = n/V=2E$.

\subsection*{3. Charged lepton field}

Assuming the same conventions for the magnetic field and considering the Landau
gauge, for the charged lepton fields we have
\begin{equation}
\psi_l^s(x) \ = \
\sum_{r=1,2} \sum_{n=0}^\infty \int \! \frac{ dq^2  dq^3}{(2\pi)^3 2 E_l}
\left[\,
b\left(\breve {q},r\right) \, U_l^s\left(x,\breve {q},r\right) + d\left(\breve {q},r\right)^\dagger
\,  V_l^{-s}\left(x,\breve {q},r\right)\,
\right]\, .
\label{fermionfieldBpart}
\end{equation}
Here, $s = \mbox{sign}(Q_l B)$ with $Q_l=-|e|$. As indicated in the main text, $\breve {q} =
(n,q^2,q^3)$ and $E_l = \sqrt{m^2_l + 2n B_e + (q^3)^2}$, with $B_e =|Q_l
B|$. The creation and annihilation operators satisfy
\begin{eqnarray}
\left\{b(\breve{q},r), b(\breve{q}\,',r')^\dagger\right\} &=&
\left\{d(\breve{q},r), d(\breve{q}\,',r')^\dagger\right\} =  \,
2 E_l\, (2\pi)^3\, \delta_{rr'}\, \delta_{nn'}\, \delta(q^2 - q^{\,\prime\, 2})\, \delta(q^3 - q^{\,\prime\, 3})\ ,
\nonumber \\
\left\{b(\breve{q},r), d(\breve{q}\,',r')^\dagger\right\} &=&
\left\{d(\breve{q},r)^\dagger, b(\breve{q}\,',r')^\dagger\right\}\ = \, 0\ .
\label{conferB}
\end{eqnarray}
In Eq.~(\ref{fermionfieldBpart}) we have also used the definitions
\begin{eqnarray}
U_l^s\left(x,\breve{q},r\right)  &=& \mathbb{E}^{s}_{\bar q}(x) \ u_l^s\left(\breve{q},r\right)\ , \nonumber\\
V_l^{-s}\left(x,\breve{q},r\right)  &=& \mathbb{\tilde{E}}^{-s}_{\bar q}(x) \
v_l^{-s}\left(\breve{q},r\right)\ ,
\label{UVlept}
\end{eqnarray}
where $\bar q = (q^0, \breve{q})$, with $q^0 = E_l$. The spinors $u_l^s$ and
$v_l^{-s}$ are given in the Weyl basis by
\begin{eqnarray}
u_l^s\left(\breve {q},r\right) &=&
\frac{1}{\sqrt{2(E_l+m_l)}}
\left( \begin{array}{c}
  ( E_l + m_l + s \sqrt{2 n B_e}\ \tau_2 - q^3 \tau_3) \phi^{(r)} \\
  ( E_l + m_l - s \sqrt{2 n B_e}\ \tau_2 + q^3 \tau_3) \phi^{(r)}\\
\end{array}
\right)\ ,
\label{uspindos}
\\
v_l^{-s}\left(\breve {q},r\right) &=&
\frac{1}{\sqrt{2(E_l+m_l)}}
\left( \begin{array}{c}
  ( E_l + m_l - s \sqrt{2 n B_e} \ \tau_2 - q^3 \tau_3) \tilde \phi^{(r)} \\
  -( E_l + m_l + s \sqrt{2 n B_e} \ \tau_2 + q^3 \tau_3) \tilde \phi^{(r)}\\
\end{array}
\right)\ ,
\label{vspindos}
\end{eqnarray}
where $\phi^{(1)}{}^\dagger = -\tilde \phi^{(2)}{}^\dagger = (1,0)$ and
$\phi^{(2)}{}^\dagger = \tilde \phi^{(1)}{}^\dagger = (0,1)$. They satisfy the
relations
\begin{eqnarray}
\sum\limits_{r=1,2} u_l^s(\breve{q}, r) \bar u_l^s(\breve{q}, r) &=&  \rlap/\!\hat q_s + m_l\ , \nonumber\\
\sum\limits_{r=1,2} v_l^{-s}(\breve{q}, r) \bar v_l^{-s}(\breve{q}, r) &=& \rlap/\!\hat q_{-s} -
m_l\ ,
\label{sumspinl}
\end{eqnarray}
where $\hat q_s^{\,\mu} = (E_l, 0, - s \sqrt{ 2 n B_e}, q^3)$. In
Eq.~(\ref{UVlept}), $\mathbb{E}^{s}_{\bar q}(x)$ and
$\mathbb{\tilde{E}}^{-s}_{\bar q}(x)$ are Ritus functions that satisfy the
eigenvalue equation
\begin{equation}
\rlap/\!{\cal D}^2 \ \mathbb{E}^{s}_{\bar q} (x) \ = \ - \left[ (q^0)^2 - 2 n B_e - (q^3)^2 \right] \mathbb{E}^{s}_{\bar q}(x)\ ,
\label{ecautov}
\end{equation}
where $\rlap/\!{\cal D} =  \rlap/\partial - i s B_e \, x^1 \gamma^2$. They
can be written as
\begin{equation}
\mathbb{E}^{s}_{\bar q} (x)\ = \ \sum_{\lambda=\pm} E^{s}_{\bar q,\lambda}(x)\,\Delta^\lambda\ ,
\qquad\qquad
\mathbb{\tilde{E}}^{-s}_{\bar q}(x) \ = \ \sum_{\lambda=\pm} E^{-s}_{\bar q,-\lambda}(x)^\ast
\Delta^{\lambda}\ ,
\label{ep}
\end{equation}
where $\Delta^{\pm} = (1\pm i \gamma^1 \gamma^2)/2$, and
\begin{equation}
E^{s}_{\bar q,\lambda}(x)  \ = \ \mathbb{F}_{(q^0,n_{s\lambda},q^2,q^3)}^s(x)\ ,
\label{autofuncion}
\end{equation}
$\mathbb{F}_{\bar q}^s(x)$ being the function defined in Eq.~(\ref{efes}).
Here the integer index $n_{s\lambda}$ is related to the quantum number $n$ by
\begin{equation}
n_{s\pm}  \ = \ n - \frac{1\mp s}{2}\ .
\label{a5}
\end{equation}
It can be seen that the spinors in Eq.~(\ref{UVlept}) satisfy
\begin{eqnarray}
\!\!\!\!\!\! \!\!\!\!\!\!\!\!\!\!
\int\! d^3 x  \, {U^s}^\dagger(x,\breve{q}, r) U^s(x,\breve{q}', r') \! &=&\! \int\! d^3 x \, {V^{-s}}^\dagger(x,\breve{q}, r) {V^{-s}}(x,\breve{q}', r') =
2 E_l  \, (2\pi)^3 \delta_{\breve{q},\breve{q}^{\prime}} \, \delta_{rr'}
\\
\!\!\!\!\!\! \!\!\!\!\!\!\!\! \!\!
\int\! d^3 x  \, {U^s}^\dagger(x,\breve{q}, r) V^{-s}(x,\breve{q}', r') \! &=& \! \int\! d^3 x \, {V^{-s}}^\dagger(x,\breve{q}, r) {U^s}(x,\breve{q}', r') =
0 \\
\!\!\!\!\!\! \!\!\!\!\!\!\!\! \!\!
\int\! d^3 x  \, \bar U^s(x,\breve{q}, r) U^s(x,\breve{q}', r') \! &=&\! -\! \int\! d^3 x \, \bar V^{-s}(x,\breve{q}, r) {V^{-s}}(x,\breve{q}', r') =  2 m_l  \, (2\pi)^3
\delta_{\breve{q},\breve{q}^{\prime}} \, \delta_{rr'}
\\
\!\!\!\!\!\! \!\!\!\!\!\!\!\! \!\!
\int\! d^3 x  \, \bar U^s(x,\breve{q}, r) V^{-s}(\tilde x,\breve{q}', r')  \!&=&\!
\int\! d^3 x \, \bar V^{-s}(x,\breve{q}, r) U^s(\tilde x,\breve{q}', r') = 0\ ,
\label{cap}
\end{eqnarray}
where $\delta_{\breve{q},\breve{q}^{\prime}}=\delta_{nn'}\delta\left(q^2-q^{\prime 2}\right)\delta\left(q^3-q^{\prime 3}\right)$
and $\tilde x = (x^0,- \vec x)$.

Following the same steps that led to Eq.~(\ref{numberb}), it is found that
the number of particles in an infinite cylinder of section $S$ lying along
the $x^1$ axis is given by $n_l = 2 E_l\, 2\pi S$.

\section*{Appendix B: Discrete symmetries}

\addtocounter{section}{1}
\setcounter{equation}{0}
\renewcommand{\theequation}{B\arabic{equation}}

In the Landau gauge, the electromagnetic interaction term between
the light quarks and the external field (chosen to be orientated
along the $z$ axis) is given by
\begin{equation}
{\cal L}(x) \ = \ -\,\sum_f \; Q_f\, B\, x^1 \,\bar \psi_f(x)\gamma_2\psi_f(x)\ ,
\label{lagext}
\end{equation}
where the sum extends over $f=u,d$, and $Q_f$ are the
corresponding electric charges. It is easy to see that the action
is separately invariant under ${\cal P}$, ${\cal C T}$ and ${\cal
PCT}$, where ${\cal P}$, ${\cal C}$ and ${\cal T}$ stand for
parity, charge conjugation and time reversal transformations
acting on the quark fields. Moreover, it can be seen that the
Lagrangian density in Eq.~(\ref{lagext}) is invariant under the
transformation ${\cal C R}_1$, where ${\cal R}_1$ is a spatial
rotation by angle $\pi$ about the $x$ axis
(i.e., a rotation that inverts the orientation of the magnetic
field $\vec B$).

The existence of these symmetries imposes constraints on the form factors in
the pion-to-vacuum hadronic matrix elements discussed in our work. As in the
case of no external field, parity is responsible for selecting which Lorentz
structures in Eq.~(\ref{vecop}) contribute to the matrix elements of the
vector and axial-vector currents, as quoted in Eqs.~(\ref{fv0}, \ref{fa0},
\ref{hvch}, \ref{hva}). Moreover, it is possible to use ${\cal CT}$ and
${\cal C R}_1$ symmetries to show that the form factors are real and equal
for both charged pions.

We start by using ${\cal CT}$ symmetry to show that the form factor
$f_{\pi^0}^{(V)}$ in Eq.~(\ref{fv01}) is real. One has
\begin{eqnarray}
\langle 0|\bar\psi_f(x)\gamma^\mu \psi_f(x)|\pi^0(\vec p\,)\rangle
& = & \langle 0|({\cal CT})^\dagger {\cal CT} \bar\psi_f(x)\gamma^\mu \psi_f(x)
({\cal CT})^\dagger {\cal CT} |\pi^0(\vec p\,)\rangle \nonumber \\
& = & \eta_T \langle 0|{\cal C}^\dagger {\cal C} \bar\psi_f(-\tilde x)
\gamma_\mu \psi_f(-\tilde x){\cal C}^\dagger {\cal C} |\pi^0(-\vec p\,)\rangle^\ast
\nonumber \\
& = & \langle 0|\bar\psi_f(-\tilde x) \gamma_\mu \psi_f(-\tilde x)
|\pi^0(-\vec p\,)\rangle^\ast \ ,
\label{ct0}
\end{eqnarray}
where $\tilde x^\mu = (x^0,-\vec x)$, and the phase $\eta_T$, arising from
the action of the time reversal operator on the pion state, has been taken
to be equal to $-1$ due to ${\cal PCT}$ invariance. From Eqs.~(\ref{fv01})
and (\ref{ct0}) it can be easily seen that $f_{\pi^0}^{(V)}$ is real. For example, from
the definition of $H^{0,\mu}_{V}(x,\vec p\,)$, Eq.~(\ref{ct0}) implies
$H^{0,0}_{V}(x,\vec p\,) \ = H^{0,0}_{V}(-\tilde x,-\vec p\,)^\ast$, which according
to the relations in Eqs.~(\ref{fv01}) leads to
\begin{equation}
- i f_{\pi^0}^{(V)} \  p^3 \ e^{-ip\cdot x} \ = \
\Big(-i f_{\pi^0}^{(V)}\Big)^\ast \ (-p^3) \ e^{i\tilde p\,\cdot(-\tilde x)}\ ,
\end{equation}
i.e.~$f_{\pi^0}^{(V)}= {f_{\pi^0}^{(V)}}^{\,\ast}$.

In the case of the matrix elements of the
axial vector current, a similar analysis leads to $f_{\pi^0}^{(Ai)} =
{f_{\pi^0}^{(Ai)}}^{\,\ast}$, for $i=1$, 2 and 3. On the other hand, taking into
account the invariance of the action under ${\cal C R}_1$ one has
\begin{eqnarray}
H_{\perp,A}^{0,\epsilon}(x,\vec p\,)
& = & \langle 0|({\cal C R}_1)^\dagger {\cal C R}_1 \bar\psi_f(x)(\gamma^1+i\epsilon\,\gamma^2)\gamma_5 \psi_f(x)
({\cal C R}_1)^\dagger {\cal C R}_1 |\pi^0(\vec p\,)\rangle \nonumber \\
& = & \langle 0|{\cal C}^\dagger {\cal C} \bar\psi_f(x')
(\gamma^1-i\epsilon\,\gamma^2)\gamma_5 \psi_f(x'){\cal C}^\dagger {\cal C} |\pi^0(\vec p\,')\rangle
\nonumber \\
& = & \langle 0|\bar\psi_f(x')
(\gamma^1-i\epsilon\,\gamma^2)\gamma_5 \psi_f(x')|\pi^0(\vec p\,')\rangle =
H_{\perp,A}^{0,-\epsilon}(x',\vec p\,') \ ,
\label{cr10}
\end{eqnarray}
where ${x'}^\mu = (x^0,x^1,-x^2,-x^3)$ and $\vec p\,' = (p^1,-p^2,-p^3)$. From
Eq.~(\ref{fa02}), this leads to
\begin{eqnarray}
-i\left( f_{\pi^0}^{(A1)} - \epsilon\, f_{\pi^0}^{(A2)} - f_{\pi^0}^{(A3)} \right)(p^1+i\epsilon\,p^2)\, e^{-ip\cdot x} & = &
\nonumber \\
& & \hspace{-4cm} -i\left( f_{\pi^0}^{(A1)} + \epsilon\, f_{\pi^0}^{(A2)} - f_{\pi^0}^{(A3)} \right)
(p^{\,\prime\,1}-i\epsilon\,p^{\,\prime\, 2})\, e^{-ip'\cdot x'} \ ,
\end{eqnarray}
which implies $f_{\pi^0}^{(A2)} = 0$.

We consider next the matrix elements with charged pion initial states.
Proceeding in a similar way as in the neutral case, from Eq.~(\ref{hvch}) we get
\begin{eqnarray}
H_{V}^{\sigma,\mu} (x,\breve p)
& = & \langle 0|({\cal CT})^\dagger {\cal CT} \bar\psi(x)\gamma^\mu \tau^{-\sigma} \psi(x)
({\cal CT})^\dagger {\cal CT} |\pi^\sigma(\breve p)\rangle \nonumber \\
& = & - \langle 0|{\cal C}^\dagger {\cal C} \bar\psi(-\tilde x)
\gamma_\mu \tau^{-\sigma} \psi(-\tilde x){\cal C}^\dagger {\cal C} |\pi^\sigma(\breve p\,')\rangle^\ast
\nonumber \\
& = & \langle 0|\bar\psi(-\tilde x) \gamma_\mu \tau^\sigma \psi(-\tilde x)
|\pi^{-\sigma}(\breve p\,')\rangle^\ast = g_{\mu\nu}H_{V}^{-\sigma,\nu} (-\tilde x,\breve p\,')^\ast \ ,
\end{eqnarray}
where we have used ${\cal C} |\pi^\pm(\bar p)\rangle = |\pi^\mp(\bar p)\rangle$
and defined $\breve p\,' =(\ell,-p^2,-p^3)$. Since $\mathbb{F}^\sigma_{\bar p}(x) =
\mathbb{F}^\sigma_{\bar p\,'}(-\tilde x)^\ast$, taking $\mu=0$ and $\sigma=-$ one obtains [see Eq.~\eqref{hvpar}]
\begin{equation}
- i f_{\pi^-}^{(V)}  \, p^3 \, \mathbb{F}^-_{\bar p}(x) \ = \
\Big(- i f_{\pi^+}^{(V)} \, p\,'^3 \, \mathbb{F}^+_{\bar p\,'}(-\tilde x)\Big)^\ast \ = \
- i \,{f_{\pi^+}^{(V)}}^{\,\ast} \, p^3\, \mathbb{F}^-_{\bar p}(x)\ ,
\end{equation}
which leads to the relation ${f_{\pi^+}^{(V)}}^{\,\ast} = f_{\pi^-}^{(V)}$. Here, $\bar p\,'=(E_{\pi^-},\breve{p}\,')$. Now, from the
invariance of the action under ${\cal C R}_1$ one has
\begin{eqnarray}
H_{V}^{\sigma,0} (x,\breve p)
& = & \langle 0|({\cal C R}_1)^\dagger {\cal C R}_1 \bar\psi(x)\gamma^0 \tau^{-\sigma} \psi(x)
({\cal C R}_1)^\dagger {\cal C R}_1 |\pi^\sigma(\breve p)\rangle \nonumber \\
& = & \langle 0|{\cal C}^\dagger {\cal C} \bar\psi(x')
\gamma^0 \tau^{-\sigma} \psi(x'){\cal C}^\dagger {\cal C} |\pi^\sigma(\breve p\,')\rangle
\nonumber \\
& = & - \langle 0|\bar\psi(x') \gamma^0 \tau^\sigma \psi(x')
|\pi^{-\sigma}(\breve p\,')\rangle = - H_{V}^{-\sigma,0} (x',\breve p\,')  \ .
\label{hchv}
\end{eqnarray}
Since $\mathbb{F}^+_{\bar p}(x) = \mathbb{F}^-_{\bar p\,'}(x')$, taking
$\mu=0$ and $\sigma=-$ one obtains $f_{\pi^-}^{(V)} = f_{\pi^+}^{(V)}$, and
then Im$\big(f_{\pi^\sigma}^{(V)}\big)= 0$.

For the matrix elements of the axial vector current, the analysis of the
zeroth and third components of the pion-to-vacuum amplitude leads to
$f_{\pi^-}^{(A1)} = f_{\pi^+}^{(A1)}$ and Im$\big(f_{\pi^\sigma}^{(A1)}\big)= 0$, while to constrain
the form factors $f_{\pi^\sigma}^{(A2)}$ and $f_{\pi^\sigma}^{(A3)}$ one needs to study
the first and second components. Taking into account the invariance under
${\cal CT}$ one has ($\epsilon=\pm$)
\begin{eqnarray}
H_{\perp,A}^{\sigma,\epsilon}(x,\breve p)
& = & \langle 0|({\cal CT})^\dagger {\cal CT} \bar\psi(x)(\gamma^1 + i\epsilon \gamma^2)\gamma_5 \tau^{-\sigma} \psi(x)
({\cal CT})^\dagger {\cal CT} |\pi^\sigma(\breve p)\rangle \nonumber \\
& = & - \langle 0|{\cal C}^\dagger {\cal C} \bar\psi(-\tilde x) (\gamma_1
- i\epsilon\gamma_2)\gamma_5 \tau^{-\sigma}\psi(-\tilde x){\cal C}^\dagger {\cal C} |\pi^\sigma(\breve p\,')\rangle^\ast
\nonumber \\
& = & + \langle 0|\bar\psi(-\tilde x) (\gamma^1 - i\epsilon\gamma^2)\gamma_5 \tau^\sigma \psi(-\tilde x)
|\pi^{-\sigma}(\breve p\,')\rangle^\ast \nonumber \\
& = & \Big( H_{A}^{-\sigma,1}(-\tilde x,\bar p\,') - i\epsilon\, H_{A}^{-\sigma,2}(-\tilde x,\breve
p\,')\Big)^\ast = \Big( H_{\perp,A}^{-\sigma,-\epsilon}(-\tilde x,\breve p\,')\Big)^\ast \ .
\label{ctch}
\end{eqnarray}
In this way, taking $\sigma=-$ in Eqs.~(\ref{hav}) and (\ref{ctch}) one obtains
\begin{equation}
\big(f_{\pi^-}^{\left(A1\right)}  + \epsilon f_{\pi^-}^{\left(A2\right)} -f_{\pi^-}^{\left(A3\right)} \big)
\; \mathbb{F}^-_{\bar p+\epsilon}(x)\ = \
\big(f_{\pi^+}^{\left(A1\right)}  + \epsilon f_{\pi^+}^{\left(A2\right)} -f_{\pi^+}^{\left(A3\right)}
\big)^\ast
\; {\mathbb{F}^+_{\bar p\,'+ \epsilon}(-\tilde x)}^{\,\ast}\ ,
\end{equation}
which implies $f_{\pi^-}^{\left(A2\right)} = {f_{\pi^+}^{\left(A2\right)}}^{\,\ast}$ and
$f_{\pi^-}^{\left(A3\right)} = {f_{\pi^+}^{\left(A3\right)}}^{\,\ast}$
(we have used the fact that $f_{\pi^-}^{(A1)}={f_{\pi^+}^{(A1)}}^{\,\ast}$). Finally, considering
${\cal C R}_1$ transformations, one has
\begin{eqnarray}
H_{\perp,A}^{\sigma,\epsilon}(x,\breve p)
& = & \langle 0|({\cal C R}_1)^\dagger {\cal C R}_1 \bar\psi(x)(\gamma^1 + i\epsilon\gamma^2)\gamma_5 \tau^{-\sigma} \psi(x)
({\cal C R}_1)^\dagger {\cal C R}_1 |\pi^\sigma(\breve p)\rangle \nonumber \\
& = & \langle 0|{\cal C}^\dagger {\cal C} \bar\psi(x') (\gamma^1
- i\epsilon\gamma^2)\gamma_5 \tau^{-\sigma} \psi(x'){\cal C}^\dagger {\cal C} |\pi^\sigma(\breve p\,')\rangle
\nonumber \\
& = & \langle 0|\bar\psi(x') (\gamma^1 - i\epsilon\gamma^2)\gamma_5 \tau^\sigma \psi(x')
|\pi^{-\sigma}(\breve p\,')\rangle \nonumber \\
& = & H_{A}^{-\sigma,1}(x',\breve p\,') - i\,\epsilon H_{A}^{-\sigma,2}(x',\breve p\,') =
H_{\perp,A}^{-\sigma,-\epsilon}(x',\breve p\,') \ ,
\label{crch}
\end{eqnarray}
which leads to $f_{\pi^-}^{\left(A2\right)} = f_{\pi^+}^{\left(A2\right)}$ and
$f_{\pi^-}^{(A3)} = f_{\pi^+}^{(A3)}$, together with
$\mathrm{Im}\big(f_{\pi^\sigma}^{(A2)}\big)=\mathrm{Im}\big(f_{\pi^\sigma}^{(A3)}\big)=0$.

\section*{Appendix C: Spatial integral}

\addtocounter{section}{1}
\setcounter{equation}{0}
\renewcommand{\theequation}{C\arabic{equation}}

We are interested in the calculation of the integral
\begin{equation}
I^{s,\lambda}_{\breve{p},\breve{q},\vec k}
 \ = \, \int d^4x \ \mathbb{F}_{\bar p}^s(x)\, E^s_{\bar{q},\lambda}(x)^\ast \; e^{ik\cdot
x}\ ,
\end{equation}
where $s,\lambda=\pm$. Here, $\mathbb{F}_{\bar p}^s(x)$ comes from the pion,
$e^{ik\cdot x}$ from the neutrino and $E^s_{\bar{q},\lambda}(x)$ from the
lepton. We use the notation already defined in Appendix A, namely $\bar q =
(E_l, \breve{q})$ and $\bar p = (E_{\pi^-}, \breve{p})$, with $\breve{q} =
(n,q^2,q^3)$ and $\breve{p} = (\ell,p^2,p^3)$, respectively. Since
$|Q_{\pi^\pm}B|=|Q_{l}B|=|eB| = B_e$ and we are interested in the situation where
$\mbox{sign}(Q_{\pi^\pm}B)=\mbox{sign}(Q_{l}B)=s$, after integration over
$x^0$, $x^2$ and $x^3$ we obtain
\begin{equation}
I^{s,\lambda}_{\breve{p},\breve{q},\vec k}
\ = \ (2\pi)^3 \delta (E_{\pi^-}-E_{\nu_l}-E_l) \, \delta (p^2-k^2-q^2) \, \delta
(p^3-k^3-q^3)\,
{\cal I}^{\,s,\lambda}_{\ell, n}(k_\perp,p^2,q^2)\ ,
\label{eqdeltas}
\end{equation}
with
\begin{eqnarray}
\hspace{-1cm} {\cal I}^{\,s,\lambda}_{\ell, n}(k_\perp,p^2,q^2) & = & \nonumber \\
& & \hspace{-2.5cm} N_\ell \,N_{n_{s\lambda}}
\, \int_{-\infty}^\infty dx^1 \,e^{-ik^1x^1}
D_\ell\left(\sqrt{2B_e}x^1-s\sqrt{\dfrac{2}{B_e}}\,p^2\right)
D_{n_{s\lambda}}\left(\sqrt{2B_e}x^1-s\sqrt{\dfrac{2}{B_e}}\,q^2\right)\ ,
\label{inti}
\end{eqnarray}
where, as also defined in Appendix A, $n_{s\lambda}= n + (s\lambda-1)/2$.
Changing variables according to
\begin{eqnarray}
\begin{cases}
\psi = \sqrt{2B_e}x^1-\dfrac{s}{\sqrt{2B_e}}\,(p^2+q^2) \\
\eta = \dfrac{s}{\sqrt{2B_e}}\,(p^2-q^2)
\end{cases}\ ,
\qquad
\begin{cases}
\sqrt{2B_e}x^1-s\sqrt{\dfrac{2}{B_e}}\,p^2 = \psi-\eta \\
\sqrt{2B_e}x^1-s\sqrt{\dfrac{2}{B_e}}\,q^2 = \psi + \eta
\end{cases}\ ,
\end{eqnarray}
and using $D_\ell(\psi-\eta)=(-1)^\ell D_\ell(\eta-\psi)$, we find
\begin{equation}
{\cal I}^{\,s,\lambda}_{\ell, n}(k_\perp,p^2,q^2)\ = \ N_\ell
\,N_{n_{s\lambda}} \dfrac{(-1)^\ell}{\sqrt{2B_e}}\, e^{-i\frac{s
k^1(p^2+q^2)}{2B_e}}  \int_{-\infty}^\infty d\psi
\,e^{-i\frac{k^1}{\sqrt{2B_e}}\psi} D_\ell\left(\eta-\psi \right)
D_{n_{s\lambda}}\left(\eta+\psi\right)\ .
\end{equation}
Next, we make use of the following property
\begin{eqnarray}
\hspace{-1cm}\int \! d\psi \, e^{i\gamma\psi} D_\ell(\eta-\psi) \, D_{n}(\eta+\psi) & = & \nonumber\\
&& \hspace{-4cm}\begin{cases}
(-1)^\ell \, \sqrt{2\pi} \, \ell! \, e^{-\frac{\gamma^2+\eta^2}{2}}\left(i\gamma+\eta\right)^{n-\ell} \, L_\ell^{n-\ell}(\eta^2+\gamma^2) \qquad \qquad  & \mathrm{if} \,\, n\geq \ell \\
(-1)^{n} \, \sqrt{2\pi} \, n! \, e^{-\frac{\gamma^2+\eta^2}{2}}\left(-i\gamma+\eta\right)^{\ell-n} \, L_{n}^{\ell-n}(\eta^2+\gamma^2)
 &  \mathrm{if} \,\, \ell\geq n \ \ .
\end{cases}
\label{prop1}
\end{eqnarray}
In our case, $\gamma=-k^1/\sqrt{2B_e}$. Owing to one of the delta functions
in Eq.~(\ref{eqdeltas}), one has $\eta=s k^2/\sqrt{2B_e}$ and therefore
$\eta^2+\gamma^2=k_\perp^2/2B_e$. Then,
\begin{eqnarray}
\hspace{-1cm}{\cal I}^{\,s,\lambda}_{\ell, n}(k_\perp,p^2,q^2)
&=& \,N_\ell \,N_{n_{s\lambda}} \sqrt{\dfrac{\pi}{B_e}}\,
e^{-i\, s\, k^1(p^2+q^2)/(2B_e)}\,e^{-k_\perp^2/(4B_e)} \nonumber\\
&& \hspace{-1.2cm} \times \begin{cases}
\ell! \, \left(\dfrac{-ik^1+s\, k^2}{\sqrt{2 B_e}}\right)^{n_{s\lambda}-\ell}
\!\! L_\ell^{n_{s\lambda}-\ell}\left(\dfrac{k_\perp^2}{2B_e}\right) \qquad
& \mathrm{if} \,\, n_{s\lambda}\geq \ell \\[4mm]
\, n_{s\lambda}!
\, \left(\dfrac{ik^1+s\, k^2}{\sqrt{2 B_e}}\right)^{\ell-n_{s\lambda}} \!\!
L_{n_{s\lambda}}^{\ell-n_{s\lambda}}\left(\dfrac{k_\perp^2}{2B_e}\right) &
\mathrm{if} \,\, \ell\geq n_{s\lambda}\ \ .
\end{cases}
\end{eqnarray}
Writing the explicit expression of $N_\ell$, $N_{n_{s\lambda}}$ and
canceling some terms, we finally arrive at
\begin{equation}
{\cal I}^{\,s,\lambda}_{\ell, n}(k_\perp,p^2,q^2)
\ = \ e^{-i\,s\, k^1(p^2+q^2)/(2B_e)}\;
\mathcal{G}^{\,s,\lambda}_{\ell,n}(k_\perp)\ ,
\label{Ir}
\end{equation}
where
\begin{eqnarray}
\mathcal{G}^{\,s,\lambda}_{\ell,n}(k_\perp) \ = \ 2\pi\,e^{-\frac{k_\perp^2}{4B_e}}
\begin{cases}
\sqrt{\dfrac{\ell!}{n_{s\lambda}!}} \, \left(\dfrac{-ik^1+s\, k^2}{\sqrt{2B_e}}\right)^{n_{s\lambda}-\ell}
\!\! L_\ell^{n_{s\lambda}-\ell}\left(\dfrac{k_\perp^2}{2B_e}\right) \ \ \mathrm{if} \,\, n_{s\lambda}\geq \ell \\
\sqrt{\dfrac{n_{s\lambda}!}{\ell!}} \, \left(\dfrac{-ik^1-s\, k^2}{\sqrt{2B_e}}\right)^{\ell-n_{s\lambda}}
\!\! L_{n_{s\lambda}}^{\ell-n_{s\lambda}}\left(\dfrac{k_\perp^2}{2B_e}\right) \ \ \mathrm{if} \,\, \ell\geq
n_{s\lambda}\ \ .
\end{cases}
\label{Mcal}
\end{eqnarray}

\section*{Appendix D: $B\to 0$ limit}

\addtocounter{section}{1}
\setcounter{equation}{0}
\renewcommand{\theequation}{D\arabic{equation}}

We provide here the explicit calculation of the integrals ${\cal
I}_{0,n}^{\,-,\lambda}$ appearing in Eq.~(\ref{V.11}). Let us consider the
integral in Eq.~(\ref{inti}) extended to a finite length, in the limit of
a weak external magnetic field. We find it convenient to define
\begin{eqnarray}
\mathcal{I}_{\ell,n}^{\,-,\lambda} & = &
N_{\ell}\,N_{n_{-\lambda}}\int_{-L}^{L}dx^{1}\,e^{ik^{1}x^{1}}
D_{\ell}\left(\sqrt{2B_{e}}\,x^{1}\right)D_{n_{-\lambda}}\left(\sqrt{2B_{e}}\,x^{1}+\sqrt{\dfrac{2}{B_{e}}}\,q^{2}\right)\
,
\label{idef}
\end{eqnarray}
where we have taken $p^2=0$, and $L$ is assumed to satisfy the conditions in
Eqs.~(\ref{V.2}) and (\ref{V.3}). It is worth noticing that
$N_{\ell}\,D_{\ell}(\sqrt{2B_{e}}\,x^{1})$ comes form the pion
field,
$N_{n_{-\lambda}}\,D_{n_{-\lambda}}(\sqrt{2B_{e}}\,x^{1}+\sqrt{(2/B_{e})}\,q^{2})$
comes from the lepton field, and the exponential comes from the antineutrino
field.

As discussed in Sec.~III.D, in the $B\rightarrow0$ limit one has
$\sqrt{2B_{e}}\,x^{1}\to 0$, therefore the $\ell =0$ and $\ell=1$ pion
wave functions satisfy
\begin{equation}
\begin{array}{ccl}
N_{0}D_{0}\big(\sqrt{2B_{e}}x^{1}\big) & \underset{{\scriptscriptstyle \sqrt{2B_{e}}x^{1}\rightarrow0}}{\longrightarrow} & \left(4\pi B_{e}\right)^{1/4}\,\,,\\
\\
N_{1}D_{1}\big(\sqrt{2B_{e}}x^{1}\big) & \underset{{\scriptscriptstyle \sqrt{2B_{e}}x^{1}\rightarrow0}}{\longrightarrow} &
\left(4\pi B_{e}\right)^{1/4}\frac{1}{2}\sqrt{2B_{e}}x^{1}\ \sim\ 0\ .
\end{array}
\end{equation}
For the lepton contribution, we have to analyze the behavior of the product
$N_{n_{-\lambda}}\,D_{n_{-\lambda}}(\sqrt{2B_{e}}x^{1}+\sqrt{(2/B_{e})}\,q^{2})$
in the limit of small $B$ and large $n$, keeping $nB$ finite. Using
Eqs.~(12.7.2), (12.10.35), (9.7.5) and (9.7.9) of Ref.~\cite{NIST}
we obtain
\begin{eqnarray}
\frac{\left(4\pi B_e\right)^{1/4}}{\sqrt{n_{-\lambda}!}}\,
D_{n_{-\lambda}}\left(\sqrt{2B_{e}}\,x^{1}+\sqrt{\dfrac{2}{B_{e}}}\,q^{2}\right)
& \simeq & \theta\left(1-r_{n_{-\lambda}}\right)
\left[\left(-1\right)^{n_{-\lambda}}\theta\left(-q^{2}\right)+
\theta\left(q^{2}\right)\right]\nonumber \\
 & & \hspace{-5.7cm} \times \, \frac{\sqrt{B_{e}}}{\left|\left(q^{2}\right)^{2}-B_{e}\left(2\,n_{-\lambda}+1\right)\right|^{1/4}}
\left[e^{i\left(\phi_{n_{-\lambda}}-q_{n_{-\lambda}}x^{1}\right)}+e^{-i\left(\phi_{n_{-\lambda}}-q_{n_{-\lambda}}x^{1}\right)}\right]
\ ,
\label{V.7}
\end{eqnarray}
where
\begin{eqnarray}
\phi_{n_{-\lambda}} & = &
\left[\frac{\left(2\,n_{-\lambda}+1\right)}{2}
\left(\arccos r_{n_{-\lambda}}-r_{n_{-\lambda}}\sqrt{1-r_{n_{-\lambda}}^{2}}\right)-\frac{\pi}{4}\right]\ ,
\nonumber \\
q_{n_{-\lambda}} & = & \frac{\left|q^{2}\right|}{q^{2}}\,\sqrt{B_{e}\left(2\,n_{-\lambda}+1\right)-\left(q^{2}\right)^{2}}\ ,
\nonumber \\
r_{n_{-\lambda}} & = &
\frac{\left|q^{2}\right|}{\sqrt{B_{e}\left(2\,n_{-\lambda}+1\right)}}\ .
\end{eqnarray}
Here, all quantities have a smooth behavior in the limit of small
$B,$ large $n$ and finite $nB,$ except $\phi_{n_{-\lambda}}.$

The above equations lead to
\begin{eqnarray}
\mathcal{I}_{0,n}^{\,-,\lambda} & = & 2\pi\left(4\pi B_{q}\right)^{1/4}\,
\theta\left(1-\frac{\left|q^{2}\right|}{\sqrt{B_{e}\left(2\,n_{-\lambda}+1\right)}}\right)
\left[\left(-1\right)^{n_{-\lambda}}\theta\left(-q^{2}\right)+\theta\left(q^{2}\right)\right]\nonumber \\
 &  & \times \,
 \frac{\sqrt{B_{e}}}{\left|\left(q^{2}\right)^{2}-B_{e}\left(2n_{-\lambda}+1\right)\right|^{1/4}}
 \left[e^{i\phi_{n_{-\lambda}}}\,\delta(k^{1}-q_{n_{-\lambda}})+
 e^{-i\phi_{n_{-\lambda}}}\,\delta(k^{1}+q_{n_{-\lambda}})\right] ,\
\label{V.9}
\end{eqnarray}
from which we can easily calculate the quantities $|{\cal
I}_{0,n}^{\,-,\lambda}|^2$ appearing in Eq.~(\ref{V.11}). We are also
interested in the product ${\cal I}_{0,n}^{\,-,+}\left({\cal
I}_{0,n}^{\,-,-}\right)^\ast$, which involves the divergent phases
$\phi_{n_{-}}$ and $\phi_{n_{+}}$. These appear through the finite
difference
\begin{equation}
\phi_{n_{-+}}-\phi_{n_{--}} \ \simeq \ -\arccos\frac{\left|q^{2}\right|}{\sqrt{2nB_{e}}}\ ,
\end{equation}
which leads to
\begin{equation}
e^{i\left(\phi_{n_{-+}}-\phi_{n_{--}}\right)} \ = \
\frac{\left|q^{2}\right|+i\,\bar q_{n}}{\sqrt{2nB_{e}}}\ ,
\end{equation}
with $\bar{q}_{n} \ = \ \sqrt{2nB_{e}-\left(q^{2}\right)^{2}}$. We obtain in
this way
\begin{eqnarray}
{\cal I}_{0,n}^{\,-,+}\left({\cal I}_{0,n}^{\,-,-}\right)^\ast & \simeq &
4\pi L\left(4\pi B_{e}\right)^{1/2}\,\theta\left(1-\frac{\left|q^{2}\right|}{\sqrt{2nB_{e}}}\right)
\frac{B_{e}}{\bar{q}_{n}}\;\frac{1}{\sqrt{2nB_{e}}}\nonumber \\
 & & \times \, \Big\{\left[-\theta\left(-q^{2}\right)\left(-q^{2}+i\,\bar{q}_{n}\right)+
 \theta\left(q^{2}\right)\left(q^{2}-i\,\bar{q}_{n}\right)\right]
 \delta (k^{1}+\bar{q}_{n})
 \nonumber \\
 & & +
\left[-\theta\left(-q^{2}\right)\left(-q^{2}-i\,\bar{q}_{n}\right)+
\text{\ensuremath{\theta}\ensuremath{\ensuremath{\left(q^{2}\right)\left(q^{2}+i\,\bar{q}_{n}\right)}}}\right]
\delta\left(k^{1}-\bar{q}_{n}\right)\Big\} \ .
\end{eqnarray}


\begin{thebibliography}{99}

\bibitem{Kharzeev:2012ph}
D.~E.~Kharzeev, K.~Landsteiner, A.~Schmitt and H.~U.~Yee,
Lect.\ Notes Phys.\  {\bf 871}, 1 (2013).

\bibitem{Andersen:2014xxa}
  J.~O.~Andersen, W.~R.~Naylor and A.~Tranberg,
  Rev.\ Mod.\ Phys.\  {\bf 88},  025001 (2016).

\bibitem{Miransky:2015ava}
  V.~A.~Miransky and I.~A.~Shovkovy,
  Phys.\ Rep.\  {\bf 576}, 1 (2015).

\bibitem{Grasso:2000wj}
  D.~Grasso and H.~R.~Rubinstein,
  Phys.\ Rep.\  {\bf 348}, 163 (2001).

\bibitem{HIC}
D.~E.~Kharzeev, L.~D.~McLerran and H.~J.~Warringa, Nucl.\ Phys.\ {\bf A803}, 227 (2008) ;
V. Skokov, A. Y. Illarionov, and V. Toneev, Int. J. Mod. Phys. A {\bf 24}, 5925 (2009);
V. Voronyuk, V. Toneev, W. Cassing, E. Bratkovskaya, V. Konchakovski, and S. Voloshin,
Phys. Rev. C {\bf 83}, 054911 (2011).

\bibitem{duncan}
R. C. Duncan and C. Thompson, Astrophys. J. {\bf 392}, L9 (1992); C.
Kouveliotou \textit{et al}., Nature (London) {\bf 393}, 235 (1998).


\bibitem{Gusynin:1994re}
  V.~P.~Gusynin, V.~A.~Miransky and I.~A.~Shovkovy,
  Phys.\ Rev.\ Lett.\  {\bf 73}, 3499 (1994); {\bf 76}, 1005(E) (1996).

\bibitem{Bali:2011qj}
  G.~S.~Bali, F.~Bruckmann, G.~Endr{\H o}di, Z.~Fodor, S.~D.~Katz, S.~Krieg, A.~Schafer and K.~K.~Szabo,
  J. High Energy Phys. 02 (2012) 044
 ;
  G.~S.~Bali, F.~Bruckmann, G.~Endr{\H o}di, Z.~Fodor, S.~D.~Katz and A.~Schafer,
  Phys.\ Rev.\ D {\bf 86}, 071502 (2012).
%
\bibitem{Nikishov:1964zza}
  A.~I.~Nikishov and V.~I.~Ritus,
  Zh. Eksp. Teor. Fiz. {\bf 46}, 776 (1964) [Sov. Phys. JETP {\bf 19}, 529 (1964)].

\bibitem{Nikishov:1964zz}
  A.~I.~Nikishov and V.~I.~Ritus,
  Zh. Eksp. Teor. Fiz. {\bf 46}, 1768 (1964) [Sov. Phys. JETP {\bf 19}, 1191 (1964)].  

\bibitem{Matese:1969zz}
  J.~J.~Matese and R.~F.~O'Connell,
  Phys.\ Rev.\  {\bf 180}, 1289 (1969).

\bibitem{FassioCanuto:1970wk}
  L.~Fassio-Canuto,
  Phys.\ Rev.\  {\bf 187}, 2141 (1969).

\bibitem{Agasian:2001ym}
  N.~O.~Agasian and I.~A.~Shushpanov,
  J. High Energy Phys. 10 (2001) 006.


\bibitem{Andersen:2012zc}
  J.~O.~Andersen,
  J. High Energy Phys. 10 (2012) 005.

\bibitem{Fayazbakhsh:2012vr}
  S.~Fayazbakhsh, S.~Sadeghian and N.~Sadooghi,
  Phys.\ Rev.\ D {\bf 86}, 085042 (2012).

\bibitem{Avancini:2015ady}
  S.~S.~Avancini, W.~R.~Tavares and M.~B.~Pinto,
  Phys.\ Rev.\ D {\bf 93}, 014010 (2016).

\bibitem{Zhang:2016qrl}
  R.~Zhang, W.~j.~Fu and Y.~x.~Liu,
  Eur.\ Phys.\ J.\ C {\bf 76}, 307 (2016).

\bibitem{Avancini:2016fgq}
  S.~S.~Avancini, R.~L.~S.~Farias, M.~Benghi Pinto, W.~R.~Tavares and V.~S.~Timóteo,
  Phys.\ Lett.\ B {\bf 767}, 247 (2017).

\bibitem{Mao:2017wmq}
  S.~Mao and Y.~Wang,
  Phys.\ Rev.\ D {\bf 96}, 034004 (2017).

\bibitem{GomezDumm:2017jij}
  D.~Gomez Dumm, M.~F.~Izzo Villafa\~ne and N.~N.~Scoccola,
  Phys.\ Rev.\ D {\bf 97}, 034025 (2018).

\bibitem{Andreichikov:2018wrc}
  M.~A.~Andreichikov and Y.~A.~Simonov,
  Eur. Phys. J. C {\bf 78}, 902 (2018).

\bibitem{Orlovsky:2013wjd}
  V.~D.~Orlovsky and Y.~A.~Simonov,
  J. High Energy Phys. 09 (2013) 136.

\bibitem{Andreichikov:2016ayj}
  M.~A.~Andreichikov, B.~O.~Kerbikov, E.~V.~Luschevskaya, Y.~A.~Simonov and O.~E.~Solovjeva,
  J. High Energy Phys. 05 (2017) 007.

\bibitem{Wang:2017vtn}
  Z.~Wang and P.~Zhuang,
  Phys.\ Rev.\ D {\bf 97}, 034026 (2018).

\bibitem{Liu:2018zag}
  H.~Liu, X.~Wang, L.~Yu and M.~Huang,
  Phys.\ Rev.\ D {\bf 97}, 076008 (2018).

\bibitem{Coppola:2018vkw}
  M.~Coppola, D.~Gomez Dumm and N.~N.~Scoccola,
  Phys.\ Lett.\ B {\bf 782}, 155 (2018).

\bibitem{Luschevskaya:2014lga}
  E.~V.~Luschevskaya, O.~E.~Solovjeva, O.~A.~Kochetkov and O.~V.~Teryaev,
  Nucl.\ Phys.\ {\bf B898}, 627 (2015).

\bibitem{Bali:2017ian}
  G.~S.~Bali, B.~B.~Brandt, G.~Endr{\H o}di and B.~Gl{\"a}{\ss}le,
  Phys.\ Rev.\ D {\bf 97}, 034505 (2018).

\bibitem{Fayazbakhsh:2013cha}
  S.~Fayazbakhsh and N.~Sadooghi,
  Phys.\ Rev.\ D {\bf 88}, 065030 (2013).

\bibitem{Simonov:2015xta}
  Y.~A.~Simonov,
  Yad. Fiz. {\bf 79}, 277 (2016) 
  [Phys. At. Nucl. {\bf 79}, 1071 455 (2016)].

\bibitem{Bali:2018sey}
  G.~S.~Bali, B.~B.~Brandt, G.~Endr{\H o}di and B.~Gl{\"a}{\ss}le,
  Phys.\ Rev.\ Lett.\  {\bf 121}, 072001 (2018).

\bibitem{Tanabashi:2018oca}
  M.~Tanabashi {\it et al.} (Particle Data Group),
  Phys.\ Rev.\ D {\bf 98}, 030001 (2018).

\bibitem{Dittrich:2000zu}
  W.~Dittrich and H.~Gies,
  Springer Tracts Mod.\ Phys.\  {\bf 166}, 1 (2000).

\bibitem{Wakamatsu:2017isl}
M.~Wakamatsu,Y.~Kitadono and P.-M.~Zhang,
Ann. Phys. (Amsterdam) {\bf 392}, 287 (2018).

\bibitem{Sokolov:1986nk}
  A.~A.~Sokolov, I.~M.~Ternov and C.~W.~Kilmister,
  {\it Radiation from Relativistic Electrons}
  (AIP, New York, 1986).

\bibitem{Peskin}
  M.~E.~Peskin and D.~V.~Schroeder,
  {\it An Introduction to Quantum Field Theory}
   (Westview Press, Reading, 1995).

\bibitem{NIST}
Digital library of mathematical functions, NIST,
https://dlmf.nist.gov.

\end{thebibliography}
\end{document}